ANALOG CIRCUIT SIZING USING MACHINE LEARNING BASED TRANSISTOR CIRCUIT MODEL

Thesis

Presented to

The Graduate Faculty of The University of Akron

In Partial Fulfillment

of the Requirements for the Degree

Master of Science

Alireza Bagheri Rajeoni

May 2021

ANALOG CIRCUIT SIZING USING MACHINE LEARNING BASED TRANSISTOR CIRCUIT MODEL

Alireza Bagheri Rajeoni

Thesis

Approved:

Advisor
Dr. Kye-Shin Lee

Committee Member
Dr. Mehdi Maleki Pirbazari

Committee Member
Dr. Ryan C. Toonen

Accepted:

Department Chair
Dr. Robert Veillette

Dean of the College
Dr. Craig Menzemer

Acting Dean of the Graduate School
Dr. Marnie Saunders

# ABSTRACT

In this work, a new method for designing an analog circuit for deep sub-micron CMOS fabrication processes is proposed. The proposed method leverages the regression algorithms with the transistor circuit model to size a transistor in 0.18 µm technology fast and without using simulation software. Threshold voltage, output resistance, and the product of mobility and oxide capacitance are key parameters in the transistor circuit model to size a transistor. For nano-scale transistors, however, these parameters are nonlinear with respect to electrical and physical characteristics of transistors and circuit simulator is needed to find the value of these parameters and therefore the design time increases. Regression analysis is utilized to predict values of these parameters. We demonstrate the performance of the proposed method by designing a Current Feedback Instrumentational Amplifier (CFIA). We show that the presented method accomplishes higher than 90% accuracy in predicting the desired value of W. It reduces the design time over 97% compared to conventional methods. The designed circuit using the proposed method consumes 5.76 µW power and has a Common Mode Rejection Ratio (CMRR) of 35.83 dB and it results in achieving 8.17 V/V gain.

DEDICATION

I dedicate this to my parents.

ACKNOWLEDGEMENT

To my heroes, my mother Tahereh Kazempour and my father Bijan Bagheri. Thank you for your love and support. A special thanks to my brothers, Amir and Majid for their continuing encouragement and support.

I would like to express my deepest gratitude to my advisor Dr. Kye-Shin Lee for his unwavering guidance and supervision. I'd also like to extend my gratitude to my committee members Dr. Mehdi Maleki Pirbazari and Dr. Ryan Toonen for their constructive advice and invaluable suggestions.

I like to acknowledge the assistance of my lab-mate, Masoud Nazari. You are ethical, easy-going, hardworking, and fun to work with. Also, I'd like to thank all my friends in the beautiful city of Akron who made my stay enjoyable.

# TABLE OF CONTENTS

Page

LIST OF FIGURES

Figure Page

LIST OF TABLES

Table Page

# CHAPTER I

# INTRODUCTION

## 1.1 Background

It was not until 1948 that the New York Times reported the invention of the transistor as a device which is applicable in instruments, such as radio frequency devices [1]. Soon after, the idea of integrated circuits began to unfold when, in 1952, Bernard M. Oliver filed a patent on integrating multiple transistors in one chip. At that time, just a few people knew the integrated circuits would dramatically change the world [2][3].

Decades later, as the integrated circuits advanced to smaller and more powerful and efficient circuits, the design procedure became challenging due to the trade-offs of design specifications like bandwidth and gain and consequently, the circuit design ended up getting time-consuming [4]. In addition, as the demand for customized integrated circuits increases, the market is looking for ways to reduce the design cost. As a result, IC designers need to come up with new methods to expedite the design of integrated circuits.

## 1.2 Motivation of the research

The operation of an amplifier depends on its DC bias point like drain current ($I_D$), drain-source voltage ($V_{DS}$), and gate-source voltage ($V_{GS}$). Also, a transistor must be biased properly in the saturation region to work as a linear amplifier. If the DC bias point of a transistor is set outside of the saturation region, for example, in the linear or cut-off region, the transconductance ($g_m$) and output resistance ($r_o$) of the transistor will drop significantly and it will not work as a linear amplifier anymore.

After deciding on the DC bias point, a design methodology is needed to find the transistor parameters to set the DC bias point. Conventional design methodologies are mainly based on using circuit simulator software to find the desired transistor parameters. These methods are however time consuming as one or more circuit simulations are needed to find the desired parameters. Therefore, there is a demand for a new design methodology to set the DC bias point fast and minimize the cost of the design.

The method proposed here leverages the transistor circuit model with regression to meet this demand. It uses the transistor equations for the saturation region and regression to predict the W which sets the DC bias point of the transistor without any circuit simulation. Therefore, it decreases the design time significantly.

## 1.3 Thesis organization

In Chapter II, an overview of previous design methodologies for analog integrated circuits is presented. Chapter III explores the transistor circuit model. In Chapter IV, regression algorithms are presented and applied to predict the Threshold Voltage ($V_{th}$), the product of Mobility and Oxide Capacitance ($\mu_n C_{ox}$), and $r_o$ under a given circuit condition. In Chapter V, the proposed method is used to design an analog circuit. In Chapter VI, the conclusion is presented.

# CHAPTER II

# PREVIOUS WORKS

Knowledge-based approach is one of the most widely used transistor sizing approaches for analog integrated circuits. It uses equations based on the circuit knowledge to design an analog circuit [5]. One of the most popular knowledge-based approaches for designing an analog circuit is the $g_m/I_D$ method. This scientific approach is first proposed in 1996 by Silveira et al. and is still being used to design analog integrated circuits [6]. This method is based on the relationship between transconductance $g_m$ over dc drain current $I_D$ and the normalized drain current $I_{D,norm}$ as given by

$$I_{D,norm} = I_D/W \qquad (2.1)$$

In this method, the value of $g_m/I_D$ for each transistor based on the design constraints is obtained. With the value of $g_m/I_D$, other parameters of the transistor that satisfies the design requirements can be found. $g_m/I_D$ is a useful parameter by which a designer can get information about the device operating region and performance. For instance, let us consider sizing one transistor shown in Figure 2.1. The constraints for this design are

$$V_{DS} = 0.6\ \text{V}$$

$$V_{GS} = 0.5\ \text{V}$$

$$L = 0.18\ \mu\text{m}$$

where L is the length of the transistor. We are looking for a width (W) that results in

$$I_D = 5\ \mu A$$

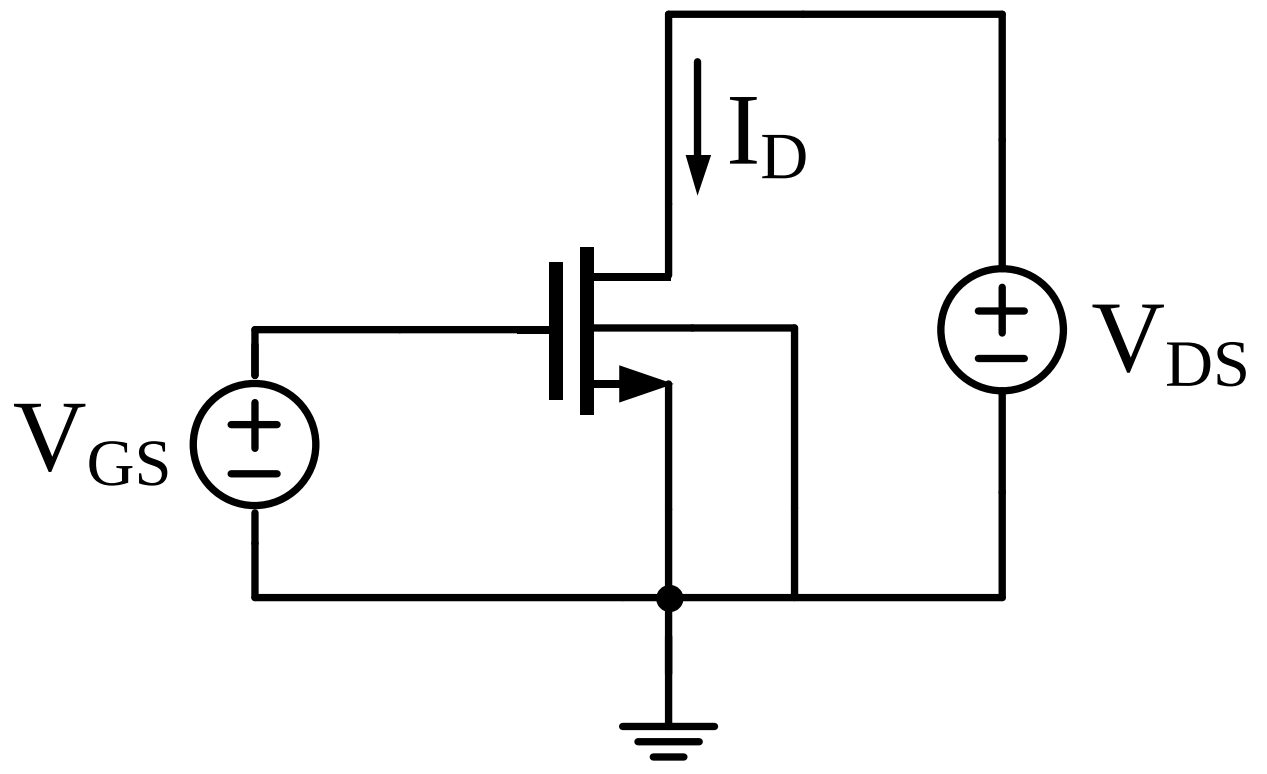

Figure 2.1 Simulation setup to size a transistor.

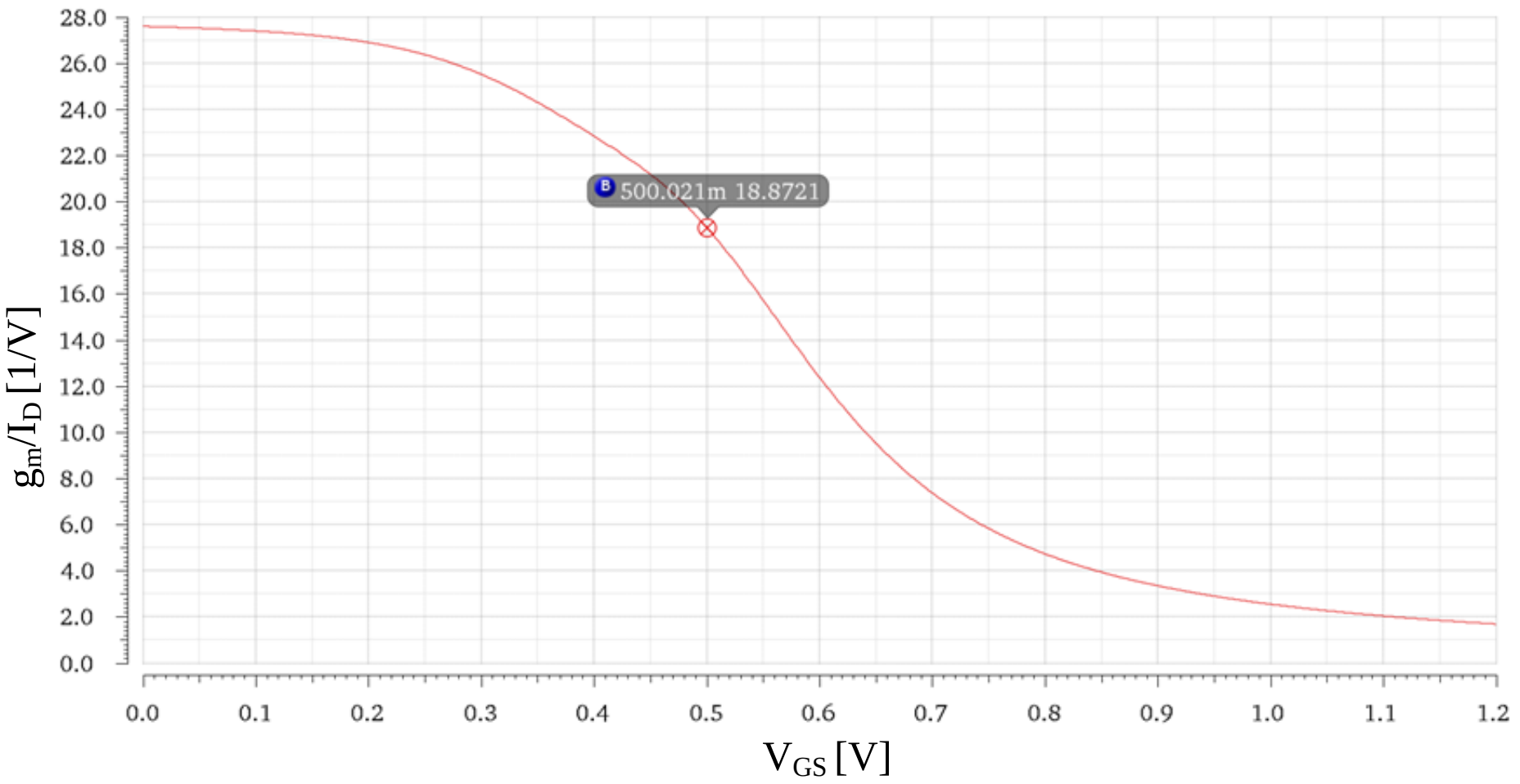

Figure 2.2 $g_m/I_D$ versus $V_{GS}$ for the example transistor.

First, we simulate $g_m/I_D$ versus $V_{GS}$ and we find the value of $g_m/I_D$ that meets the requirement for $V_{GS} = 0.5$ V. The derived $g_m/I_D$ is 18.87 $V^{-1}$ based on Figure 2.2. Then we simulate $g_m/I_D$ versus normalized current ($I_D/W$) to find the value of W that meets the constraint of $g_m/I_D = 18.87\ V^{-1}$ as shown in Figure 2.3. W can be calculated as

$$\frac{I_D}{W} = 3.393\ A/m, W = \frac{5\ \mu A}{3.393\ A/m} = 1.47\ \mu m \tag{2.2}$$

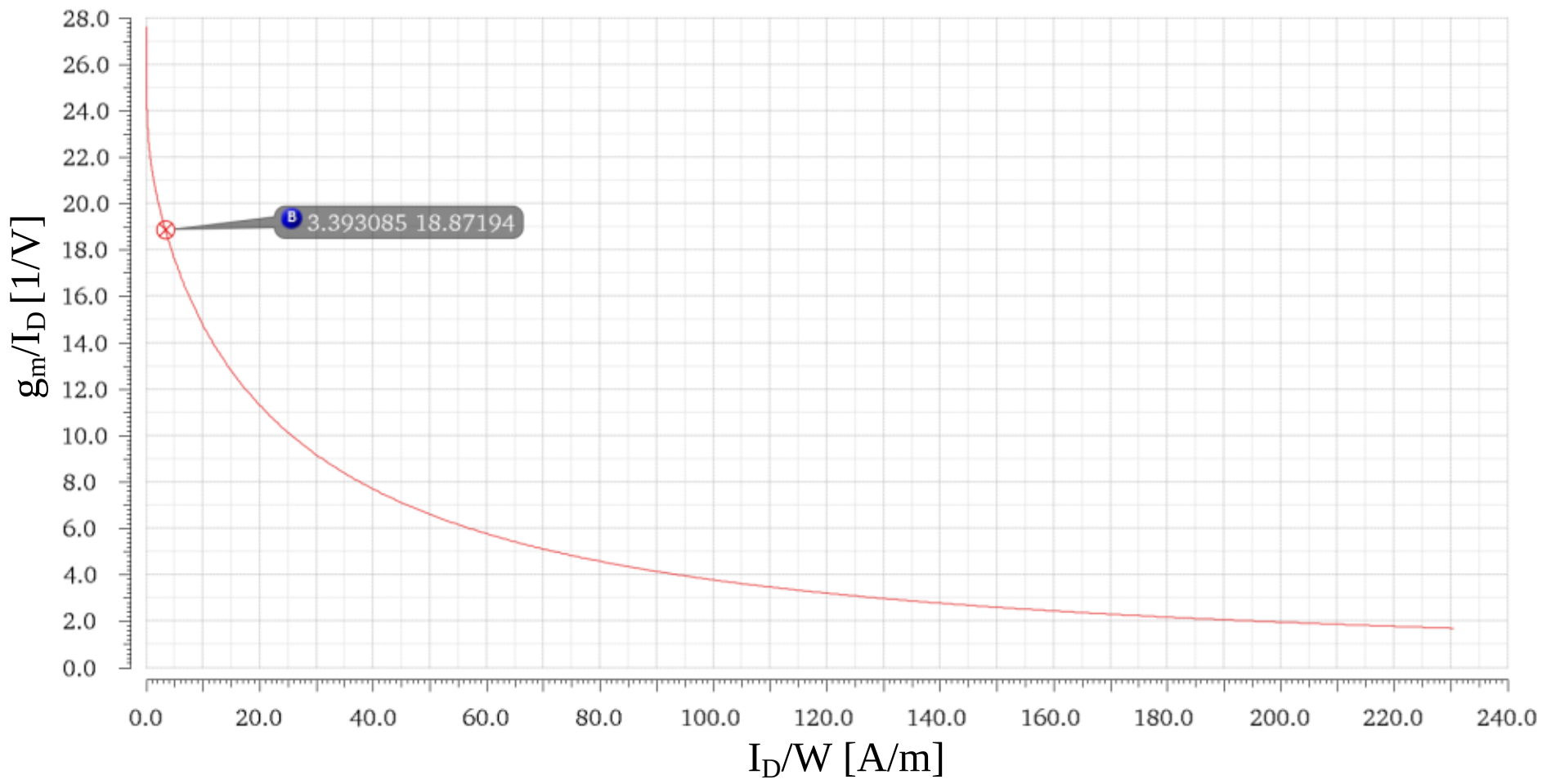

Figure 2.3 $g_m/I_D$ versus $I_D/W$ plot for the example transistor.

$I_D/W$ versus $V_{GS}$ can also be used to directly find the W that meets the design constraints. It results in $W = 1.47\ \mu m$ as well.

If we use the derived W from the $g_m/I_D$ method alongside the design constraints and simulate circuit shown in Figure 2.1, we will obtain $I_D = 4.54\ \mu A$ which is lower than the desired $I_D$. This means the $g_m/I_D$ method needs adjustment to find the correct value of W. Also, this method uses a circuit simulator to find the desired parameter which increases the time of the design. Please note, a linear sweep with 1000 steps in Cadence Spectra was used to obtain these figures.

Another conventional and widely used method for the transistor sizing is to use a brute-force approach. Brute-force is a simulation-based method in which a designer sweeps one design parameter while fixing all the other design parameters until the desired outcome is achieved. However, sometimes two or more design variables need to be tweaked to satisfy the design constraints. This method simulates each transistor separately to find the value of a parameter that results in the desired DC bias point. For instance, it sets $V_{DS}$, L, and $V_{GS}$ of the transistor and finds the desired W under a fixed $I_D$. However, this method is time-consuming as designer need to simulate each transistor using simulation software to find the W that sets the desired DC bias point. Indeed, the design time is high. Moreover, with the increase in the number of transistors in a circuit, the slight mismatches between the obtained DC bias point and the desired DC bias point of each transistor causes the obtained DC bias point of the circuit to deviate hugely from the desired DC bias point of the circuit. Hence, adjustment will be needed to correct the DC bias point of the circuit which increases the design time further. To illustrate this method further, let us consider the previous design problem. For this problem, we design the circuit shown in Figure 2.1 in the simulation software. Then, we sweep W until we find the W that results in $I_D = 5\ \mu A$ while keeping $V_{DS}$, $V_{GS}$, and L at 0.6 V, 0.5 V, and 0.18 µm, respectively. In Figure 2.4, we sweep W from 0.2 µm to 10 µm and output $I_D$. We can see that for $W = 1.64\ \mu m$ we can achieve $I_D = 5\ \mu A$. Please note, a linear sweep with 1000 steps in Cadence Spectra is used to obtain these figures.

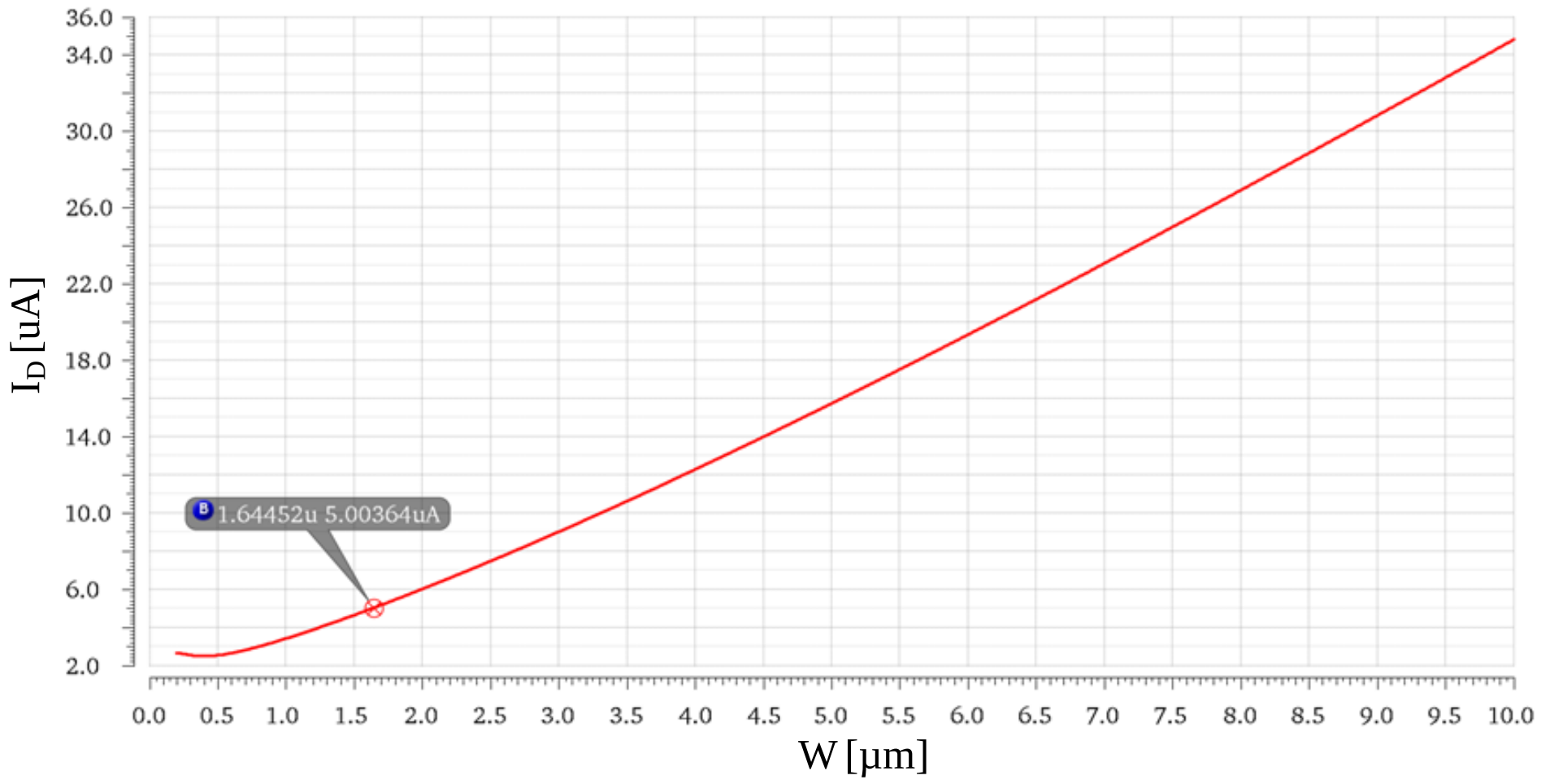

Figure 2.4 $I_D$ versus W plot for the example transistor.

Another approach for designing analog circuit is based on optimizing the design. In this approach, a set of metrics based on the design requirements and a cost function based on the metrics are formulated. Then algorithms are used to minimize the cost function. The optimization-based method addresses sizing problems through equation-based, simulation-based, and learning-based techniques [5]. [7] describes an example of a simulation-based approach. It uses genetic algorithms to randomly initiate a population and mutate the best individuals in the population for reproduction unless a better result is not achievable. This method however does not always guarantee a convergence.

So far, the learning-based method is the most promising approach within optimization-based approach to optimize a design. It is based on the machine learning algorithms to size transistors in analog circuits. In this approach, machine learning algorithms design mathematical models using the existing dataset to make predictions or decisions without being trained explicitly to do so [8]. [9] uses artificial neural networks for designing analog circuits. It trains Recurrent Neural Network (RNN) and Deep Neural

Network (DNN) to automatically size transistors and concludes that the DNN achieves a better performance. The algorithm, however, needs to be retrained for each new analog circuit. [10] leverages reward-based reinforcement learning to iteratively update the parameters of transistors and check the result to optimize the design by maximizing the reward. However, the algorithm needs to be retrained if the design specification changes. [11] leverages reinforcement learning structured with neural network to accumulate rewards to optimize sizing. It considers the post layout constraints like layout parasitics in the algorithms as well. Nonetheless, convergence is very time consuming and the codes need to be retrained for each new specification.

In this work, we will propose a new analog sizing methodology using a combination of the transistor circuit model and regression algorithms. In the next chapter, we will talk about the transistor circuit model equations. Subsequently, we present the regression algorithms to improve those equations to accurately size a transistor.

CHAPTER III

TRANSISTOR CIRCUIT MODEL

The transistor circuit model uses equations based on the physical properties of the transistor in the saturation region to set the DC operation point. The transistor model finds the aspect ratio $(W/L)$ by using threshold voltage, Gate to Source Voltage ($V_{GS}$), Drain to Source Voltage ($V_{DS}$), Oxide Capacitance ($C_{ox}$), Mobility ($\mu_n$), and output resistance. In saturation, we can model the transistor with two currents, active current and passive current as it has been shown in Figures 3.1 and 3.2. These Figures represent the transistor circuit model in the saturation region. The requirements for the saturation region for the NMOS are $V_{GS}$>$V_{th}$ and $V_{DS}$>$V_{GS} - V_{th}$ and for the PMOS are $|V_{DS}| > |V_{GS}| - |V_{DS}|$ and $|V_{GS}| > |V_{th}|$ [12]. For the NMOS transistor, we have

$$I_A = 0.5\mu_n C_{ox} \left(\frac{W}{L}\right) (V_{ov})^2 \tag{3.1}$$

$$I_P = \frac{V_{DS} - V_{ov}}{r_o} \tag{3.2}$$

$$I_D = I_A + I_P \tag{3.3}$$

where $V_{ov} = V_{GS} - V_{th}$.

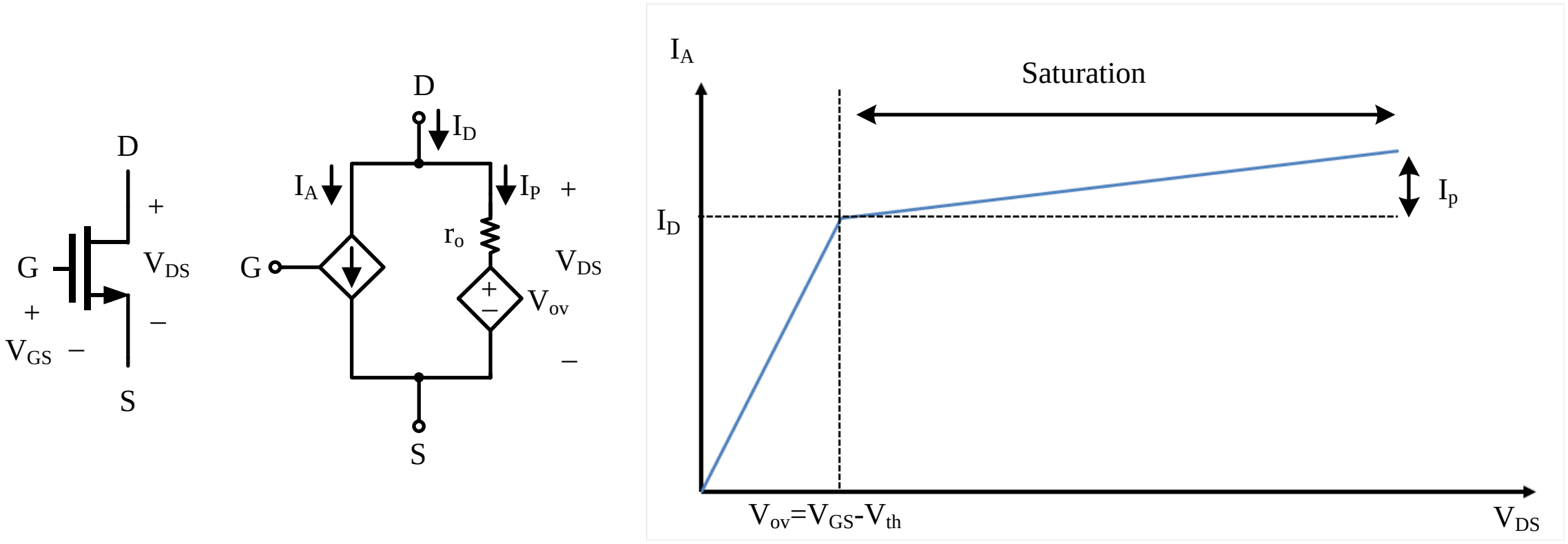

Figure 3.1 The NMOS transistor circuit model.

And for the PMOS transistor, we have

$$I_A = 0.5\mu_p C_{ox}\left(\frac{W}{L}\right)\left(V_{ovp}\right)^2 \tag{3.4}$$

$$I_P = \frac{|V_{DS}| - |V_{ovp}|}{r_o} \tag{3.5}$$

$$I_D = I_A + I_P \tag{3.6}$$

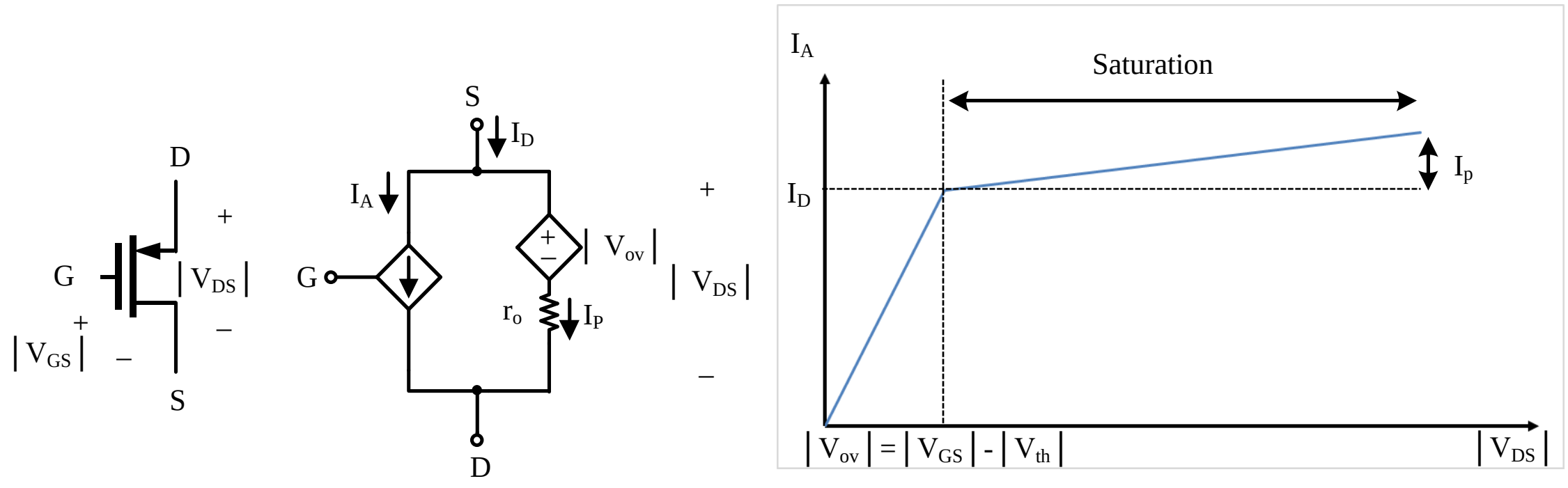

Figure 3.2 The PMOS transistor circuit model.

where $|V_{ovp}| = |V_{GS}| - |V_{th}|$.

After the designer specifies $I_D$, $V_{DS}$, and $V_{GS}$, $I_P$ and $I_A$ can be calculated using estimated or measured values of $r_o$, $V_{th}$, and $\mu_p C_{ox}$ or $\mu_n C_{ox}$. Then, $\frac{W}{L}$ ratio can be calculated using (3.1) or (3.4) equations and finally the W that meets the specification can be derived. It should be noted that the transistor circuit model approach is usually an iterative process. Thus, designers need to simulate the circuit with the initially calculated value of the W, then measure $r_o$, $V_{th}$, and $\mu_p C_{ox}$ or $\mu_n C_{ox}$ from the simulator to recalculate W. A satisfactory result can be achieved using the transistor model methodology for 0.35 μm technology or higher with just one iteration. But for technology smaller than 0.35 μm, it gets difficult to design a circuit using this approach due to the high nonlinearity in $\mu_n C_{ox}$ and $V_{th}$ in deep submicron transistors. In the next chapter, we use regression methods to estimate $\mu_n C_{ox}$ and $V_{th}$ with a function of W and L to solve this problem.

CHAPTER IV

# MODELING THE CMOS TRANSISTOR CHARACTERISTICS

Based on the experiment results, we found $\mu_n C_{ox}$ and $V_{th}$ in 0.18 μm CMOS technology highly fluctuate with respect to physical and electrical properties of the transistor. Therefore, it is almost impossible to design an analog circuit in 0.18 μm technology using iterative design process of the transistor circuit model.

In this chapter, we leverage regression algorithms to estimate a model to predict $\mu_n C_{ox}$ and $V_{th}$ in design process. First, an overview of the regression algorithms for approximating the relationship between a dependent variable and one or more independent variables is presented. Then we show the procedure for deriving the regression expressions to predict $\mu_n C_{ox}$ and $V_{th}$ in 0.18 μm technology. For simplicity, we only derive regression models for $\mu_n C_{ox}$ and $V_{th}$ with respect to W and L, and fix the $V_{GS}, V_{SB}$, and $V_{DS}$. Finally, to fully automate the design process, we present a first order model to estimate $r_o$ with respect to L.

## 4.1 Regression

In this section, we introduce linear and nonlinear regression models as machine learning techniques for modeling the relationship between the $\mu_n C_{ox}$ and $V_{th}$ with W and

L. We describe the algorithms in detail and show how these two methods find the optimum models that result in the best approximation of the dependent variable.

4.1.1 Nonlinear Regression

We used Guassian-Newton algorithms described in [13] to find an optimum function to predict nonlinear relationship between two or more variables. This Guassian-Newton algorithm is an iterative method used to find the nonlinear least square error. For a set of design vectors $x_n$, $n = 1, 2 \dots N$ and parameters $\theta = \theta_1, \theta_2, \dots \theta_p$, we are looking for an expected function $f(x_n, \theta)$ to approximate data vector y which we have obtained from the circuit simulation. Taylor series of $f(x_{n,}\theta)$ is

$$f(x_{n,}\theta) = f(x_n, \theta^0) + v_{n1}(\theta_1 - \theta_1^0) + v_{n2}(\theta_2 - \theta_2^0) + \cdots + v_{nP}(\theta_p - \theta_p^0) \quad (4.1)$$

where $v_{np} = \frac{\partial f(x_{n,}\theta)}{\partial \theta_p} \Big|_{\theta^0}$, and $\theta^0$ depicts the value of the parameters given in the first iteration. If we define ES as the expected surface, we have

$$ES(\theta) = ES(\theta^0) + V^0(\theta - \theta^0) \quad (4.2)$$

where $V^0$ is a $N \times P$ derivative matrix with elements $\{v_{nP}\}$. If we define z as the residuals

$$z(\theta) = y - ES(\theta) = y - (ES(\theta^0) + V^0\delta) = z^0 - V^0\delta \quad (4.3)$$

where $\delta = \theta - \theta^0$ and $z^0 = y - ES(\theta^0)$. Now we need to find the guassian increment $\delta^0$ for which we minimize the approximate residual sum of squares $\|z^0 - V^0\delta\|^2$. Therefore, we use the following decomposition of the $V^0$

$$V^0 = QR = Q_1R_1 \tag{4.4}$$

where $Q^T = H_{u1}H_{u2} \ldots H_{up}$, $H_u = I^T - 2uu^T$ is a Householder transformation, I is a $N \times N$ identity matrix, u is a N-dimentional unit vector and R is a $N \times P$ matrix and Q is orthogonal ( $Q^TQ = QQ^T = I$ ) [14]. If we transform the residuals vector to $w_1$

$$w_1 = Q_1^T z^0 \tag{4.5}$$

then the projection of w to the expected surface will be

$$ES = Q_1w_1 \tag{4.6}$$

since $ES = x_n\delta^0$, by using the decomposition of the $x_n$ we have

$$Q_1R_1\delta^0 = Q_1w_1 \tag{4.7}$$

$$R_1\delta^0 = w_1 \tag{4.8}$$

therefore, $ES(\theta^1) = ES(\theta^0 + \delta^0)$ should be closer to y than $ES(\theta^0)$, and the algorithm perform this process iteratively to minimize the value of the residual sum of squares.

### 4.1.2 Linear Regression

We used the least square error technique presented in [13] to derive a function to predict the linear relationship between one or more variables. For a dependent variable y and one or more independet variables X, suppose we want to find the following model

$$y = \beta_0 + \beta_1x_1 + \beta_2x_2 + \cdots + \beta_nx_n \tag{4.9}$$

Therefore, we need to find β that gives us the minimum error. From simulation, we have observations $Y = \begin{bmatrix} y_1 \\ \vdots \\ y_N \end{bmatrix}$, and the independent data matrix $X_{n\times(p+1)}$ can be written as

$$X = \begin{bmatrix} 1 & x_{11} & \dots & x_{1p} \\ \vdots & \vdots & \ddots & \dots \\ 1 & x_{n1} & \dots & x_{np} \end{bmatrix} \tag{4.10}$$

Therefore, the expectation function is

$$Y' = X\,B \tag{4.11}$$

where $B = \begin{bmatrix} \beta_0 \\ \vdots \\ \beta_p \end{bmatrix}$. Now the sum of the squared error is

$$SSE = (Y - Y')^2 = (Y - XB)^2 = \sum_{n=1}^{N} \left( y_n - \sum_{p=1}^{P+1} x_{np}\beta_p \right)^2 \tag{4.12}$$

We can minimize the SSE equation by taking its derivitive and solving it for B which will lead to optimum coefficients B

$$\frac{\partial SSE}{\partial \beta_p} = 0 \;\; => \;\; B = (X^T X)^{-1} X^T Y \tag{4.13}$$

## 4.2 Modeling the transistor characteristics using regression

In this section, we use regression techniques to approximate the relationship between $\mu_n C_{ox}$ and $V_{th}$ versus W and L with optimum functions to improve the transistor circuit modeling. First, we define general values for independent variables as shown in Table 4.1. Decreasing the dimension of W and L any further than 1 µm makes the model highly complex in 0.18 µm technology as the interaction terms between W and L appears. We choose 0.5 V for $V_{GS}$ which is greater than the threshold voltage just to make sure that the overdrive voltage is always positive especially with all the possible variation in the threshold voltage. We select $0.6\,\mathrm{V}$ for $V_{DS}$ which is below $V_{DD}/2$ and higher than the overdrive voltage.

Table 4.1 General values for independent variables.

| | W [μm] | L [μm] | $V_{GS}$ [V] | $V_{DS}$ [V] |
|---|---|---|---|---|
| General Values | 1 | 1 | 0.5 | 0.6 |

4.2.1 NMOS

To model the NMOS characteristic, first we need to estimate the $\mu_n C_{ox}$, $V_{th}$, and $r_o$ while changing their independent variabls. We use the setup shown in Figure 4.1 to obtain plots that show the relation between $\mu_n C_{ox}$ and $V_{th}$ with W and L. We use Cadence Spectra to obtain the plots.

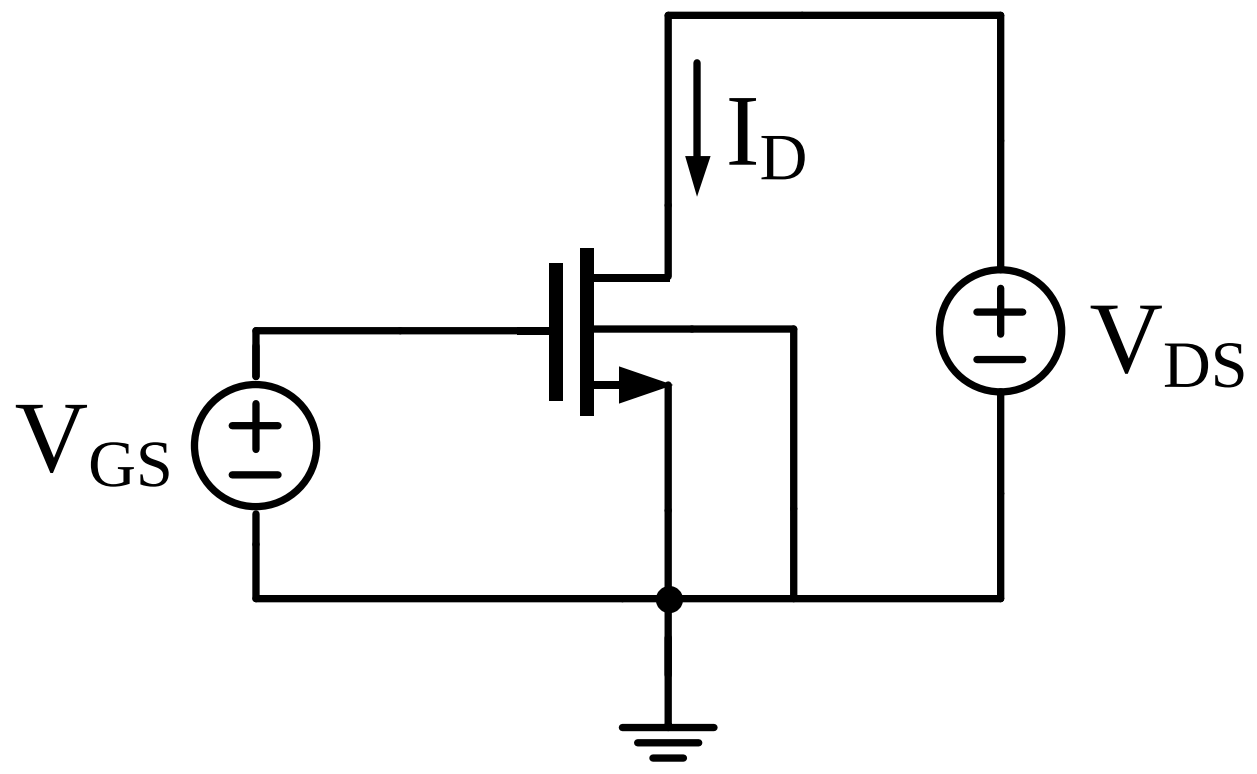

Figure 4.1 Simulation setup for obtaining NMOS $\mu_n C_{ox}$ and $V_{th}$ versus W and L plots.

The threshold voltage $V_{th}$ versus W is shown in Figure 4.2. W is swept from 1 μm to 50 μm while all other independent variables are fixed at their general values as mentioned in Table 4.1. As shown from the plot, the relationship between $V_{th}$ and W is highly nonlinear. This relationship looks like a combination of concave asymptotic and sigmoid regression models. Therefore, we use the following regression model to estimate the relation between $V_{th}$ and W

$$V_{th}(W) = \theta_1 - \theta_2 \times \exp(-\theta_3 \times W) + \theta_4 \times \exp\,(W^{\theta_5}) \tag{4.14}$$

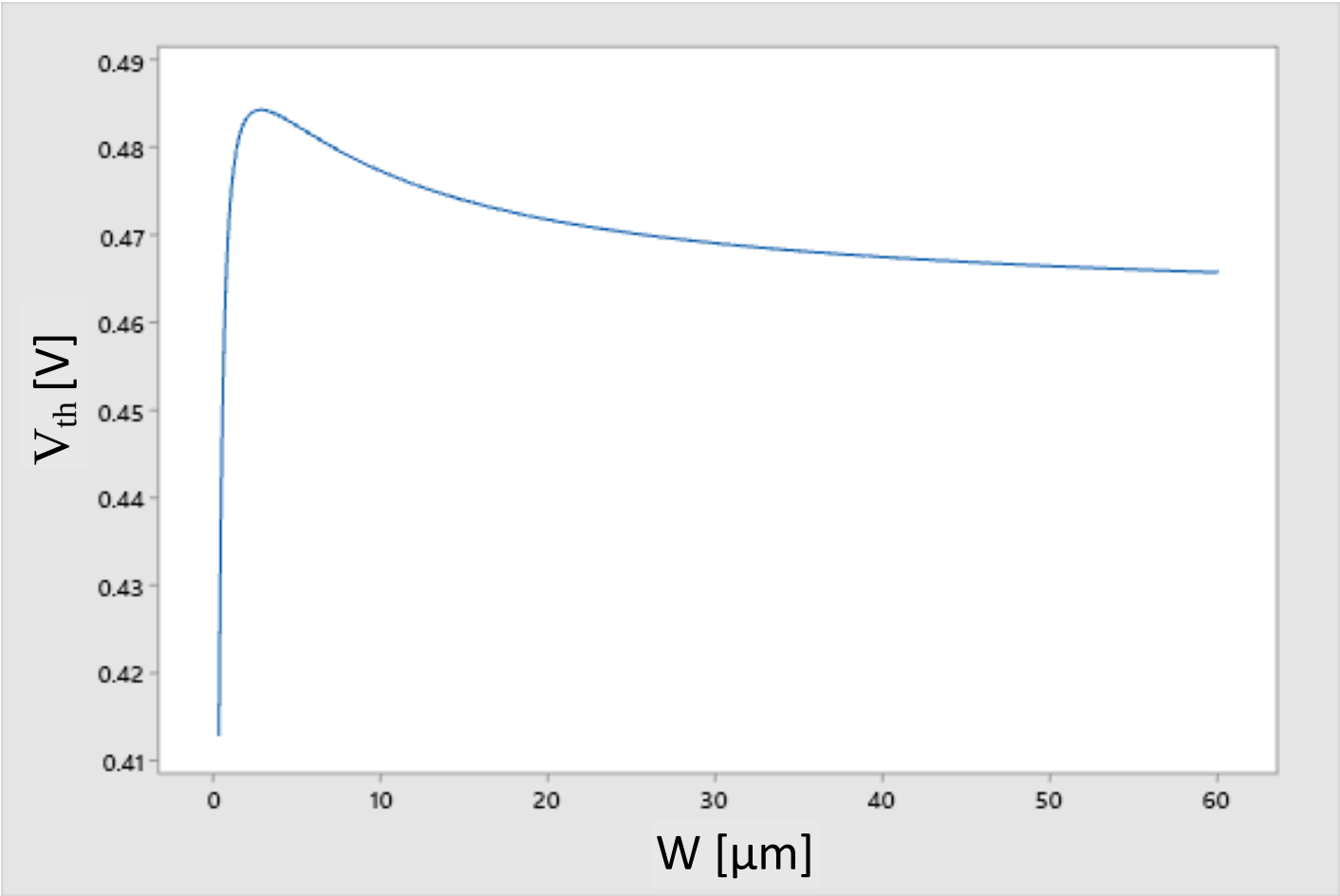

Figure 4.2 $V_{th}$ versus W plot obtained from circuit simulation while other variables are kept at general values in Table 4.1.

Now we use nonlinear regression method presented in 4.1.1 to estimate the θ's so that $V_{th}(W)$ most closely fits the $V_{th}$ versus W plot. The estimated model is

$$\begin{aligned} V_{th}(W) = 0.10471 - 0.14941 \times \exp(-2.0794 \times W \times 10^6) \\ + 0.14273 \times \exp\,((W \times 10^6)^{-0.01878}) \end{aligned} \tag{4.15}$$

The percent error between the regression and circuit simulation is shown in Figure 4.3. We can see that the percent error is less than 3% and almost zero for W higher than 20 µm.

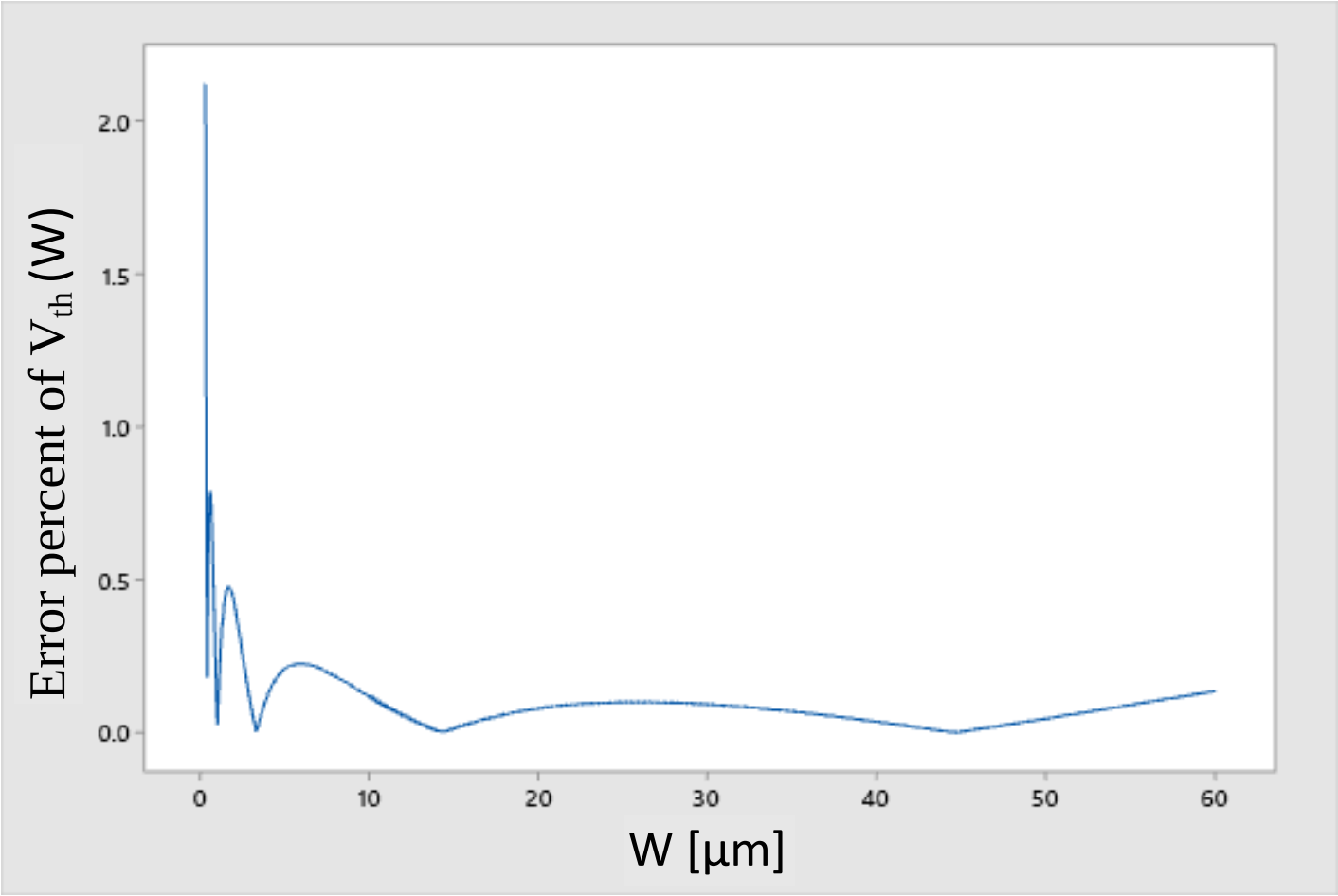

Figure 4.3 Error percent for $V_{th}$ versus W between simulation and regression.

Figure 4.4 shows $V_{th}$ versus L. In this relationship, the algorithm only converges for a combination of convex asymptotic regression and exponential regression models. The regression model is

$$V_{th}(L) = -\theta_1 \times L^{-\theta_2} + \theta_3 \times \exp\left(L^{-\theta_4}\right) \tag{4.16}$$

After using the nonlinear regression method presented in 4.1.1, the estimated model is

$$\begin{aligned} V_{th}(L) = {} & -0.42488 \times (L \times 10^6)^{-0.27216} \\ & + 0.28097 \times \exp\left((L \times 10^6)^{-0.20575}\right) \end{aligned} \tag{4.17}$$

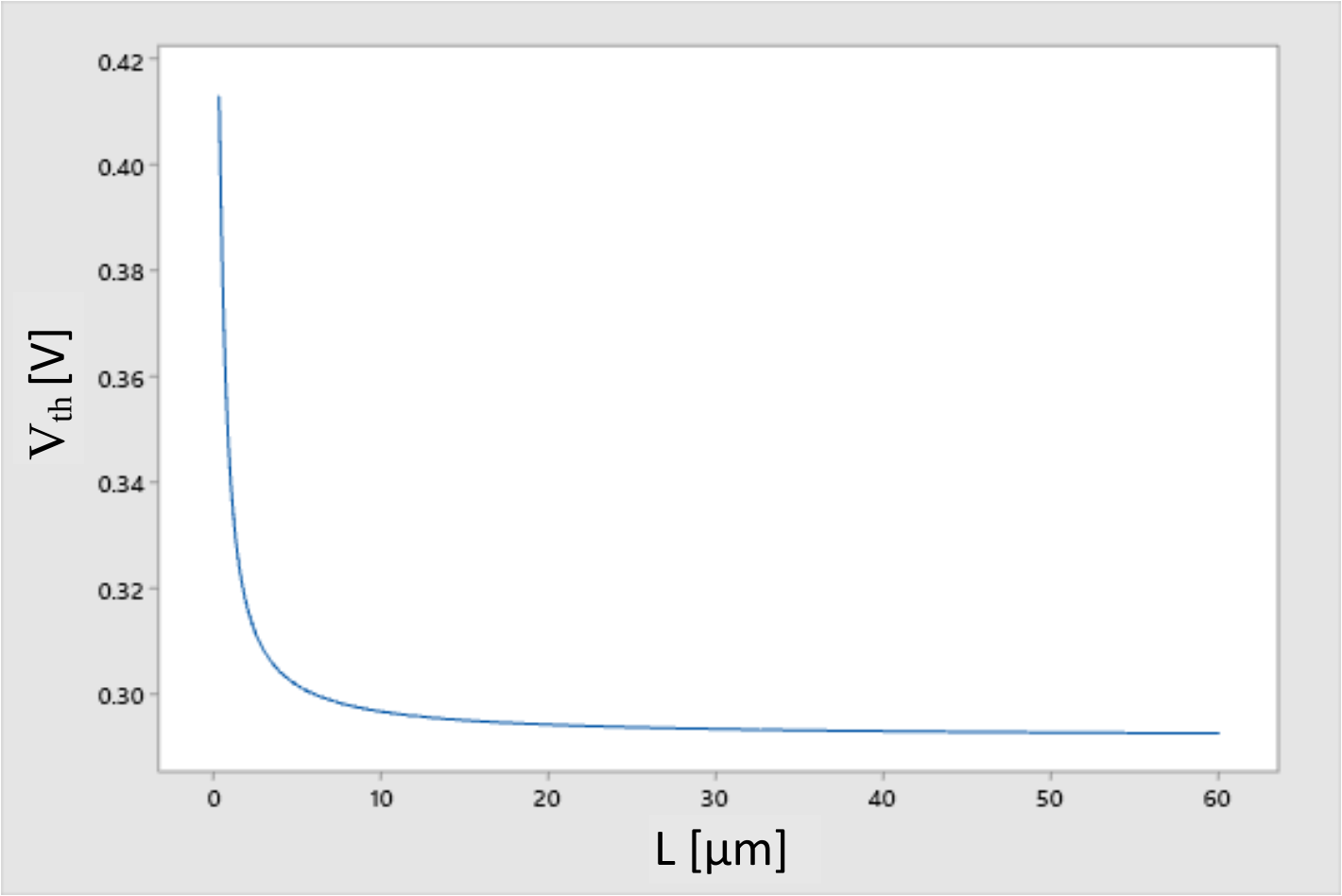

Figure 4.4 Simulated $V_{th}$ versus L plot obtained from circuit simulation while other variables are kept at general values in Table 4.1.

The error percent for the estimated model is shown in Figure 4.5. The error percent is almost zero for L larger than 5 µm.

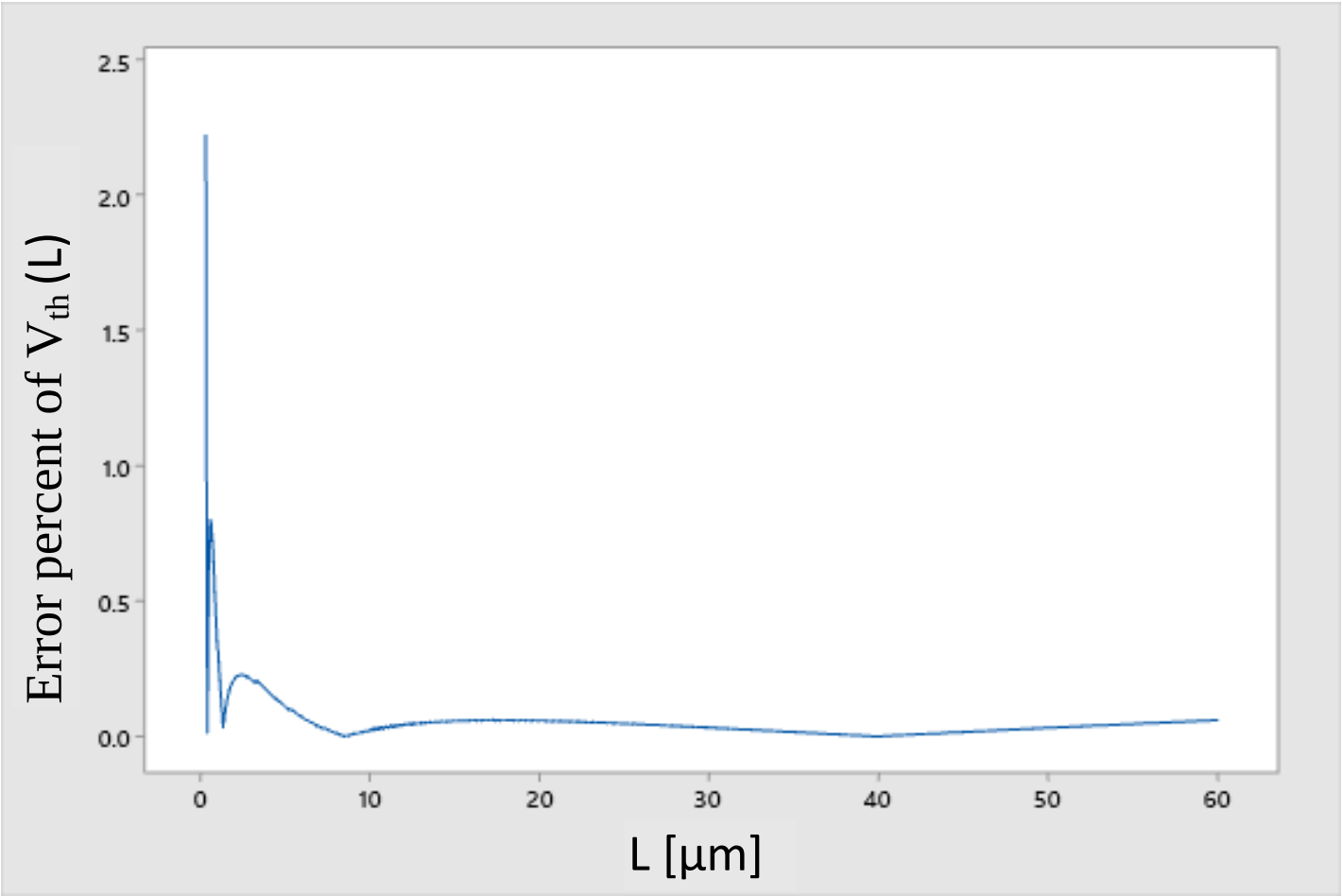

Figure 4.5 Error percent for $V_{th}$ versus L between simulation and regression.

After finding the individual relationships for $V_{th}$ with W and L, we need to find how W and L jointly change $V_{th}$. We pick thirteen points starting from 0.3 µm to 10 µm for W and L since this range covers the most variation in $V_{th}$ based on Figures 4.2 and 4.4. Then we simulate $V_{th}$ versus W and L for these points. We use the following regression model to estimate the relationship.

$$V_{th}(W, L) = \beta_0 + \beta_1 V_{th}(W) + \beta_2 V_{th}(L) \tag{4.4}$$

The linear regression method presented in 4.1.2 is used to estimate β values so that $V_{th}(W, L)$ most closely fits the value we derived from the simulation. The final estimated model for the relation for $V_{th}$ with W and L is

$$\begin{aligned} V_{th}(W, L) = -0.38164 + 0.9422 \times V_{th}(W) \\ + 0.98848 \times V_{th}(L) \end{aligned} \tag{4.5}$$

To obtain $\mu_n C_{ox}$ versus W and L plots using circuit simulator, we need to define $\mu_n C_{ox}$ as an expression since there is no option in Cadence to directly output it. We use the following expression to obtain the $\mu_n C_{ox}$ plots

$$\mu_n C_{ox} = \frac{2I_D}{\frac{W}{L}(V_{GS} - V_{th})^2(1 + \lambda V_{DS})} \tag{4.6}$$

where λ is channel length modulation coefficient. In circuit simulation, we sweep W and L and output $\mu_n C_{ox}$ using equation (4.18). Figure 4.6 shows $\mu_n C_{ox}$ versus W where W is swept from 1 μm to 50 μm while all other independent variables are fixed at their general values.

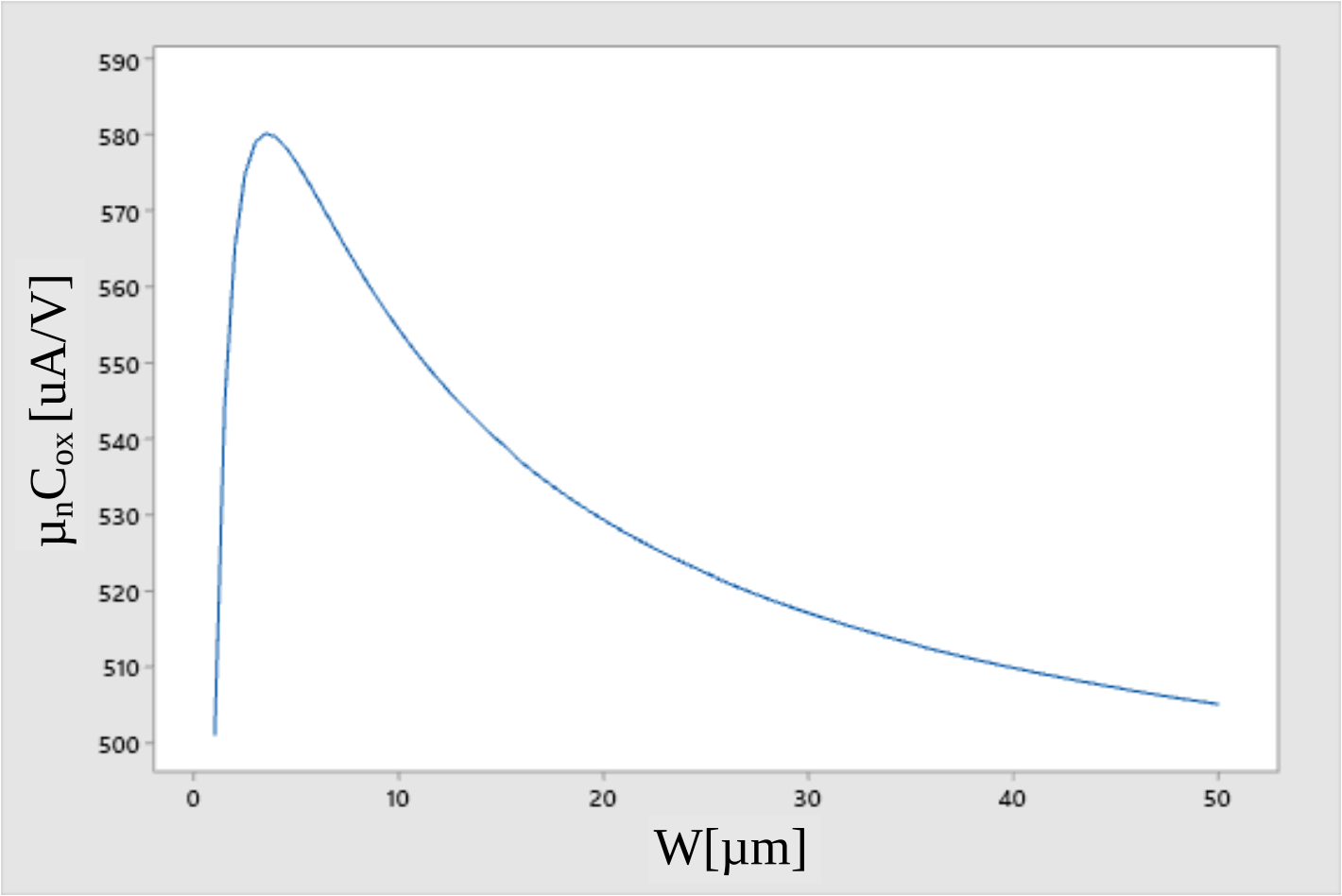

Figure 4.6 $\mu_n C_{ox}$ versus W plot obtained from circuit simulation while other variables are kept at general values in Table 4.1.

As shown in this Figure, the relation between $\mu_n C_{ox}$ and W is highly nonlinear. We use a combination of concave asymptotic and exponential regression model for this relationship

$$\mu_n C_{ox}(W) = \theta_1 - \theta_2 \times \exp(-\theta_3 \times W) + \theta_4 \times W^{-\theta_5} \qquad (4.21)$$

Now we use nonlinear regression method presented in 4.1.1 to estimate the unknown parameters θ's so that $\mu_n C_{ox}(W)$ most closely fits the $\mu_n C_{ox}$ versus W. The estimated model is as

$$
\begin{aligned}
\mu_n C_{ox}(W) = -388.11 \times 10^{-6} \\
- 428.19 \times 10^{-6} \\
\times \exp(-988351 \times W \times 10^{6}) \\
+ 273.7 \times 10^{-6} \times (W \times 10^{6})^{-0.21960}
\end{aligned}
\quad (4.22)
$$

The error percent for the estimated model for this fit is shown in Figure 4.7. We can see that the error percent is small and is limited to less than 0.3% for W larger than 5 μm. High error percent for W less than 5 μm is due to high nonlinearity of W in this region as we can see from Figure 4.6.

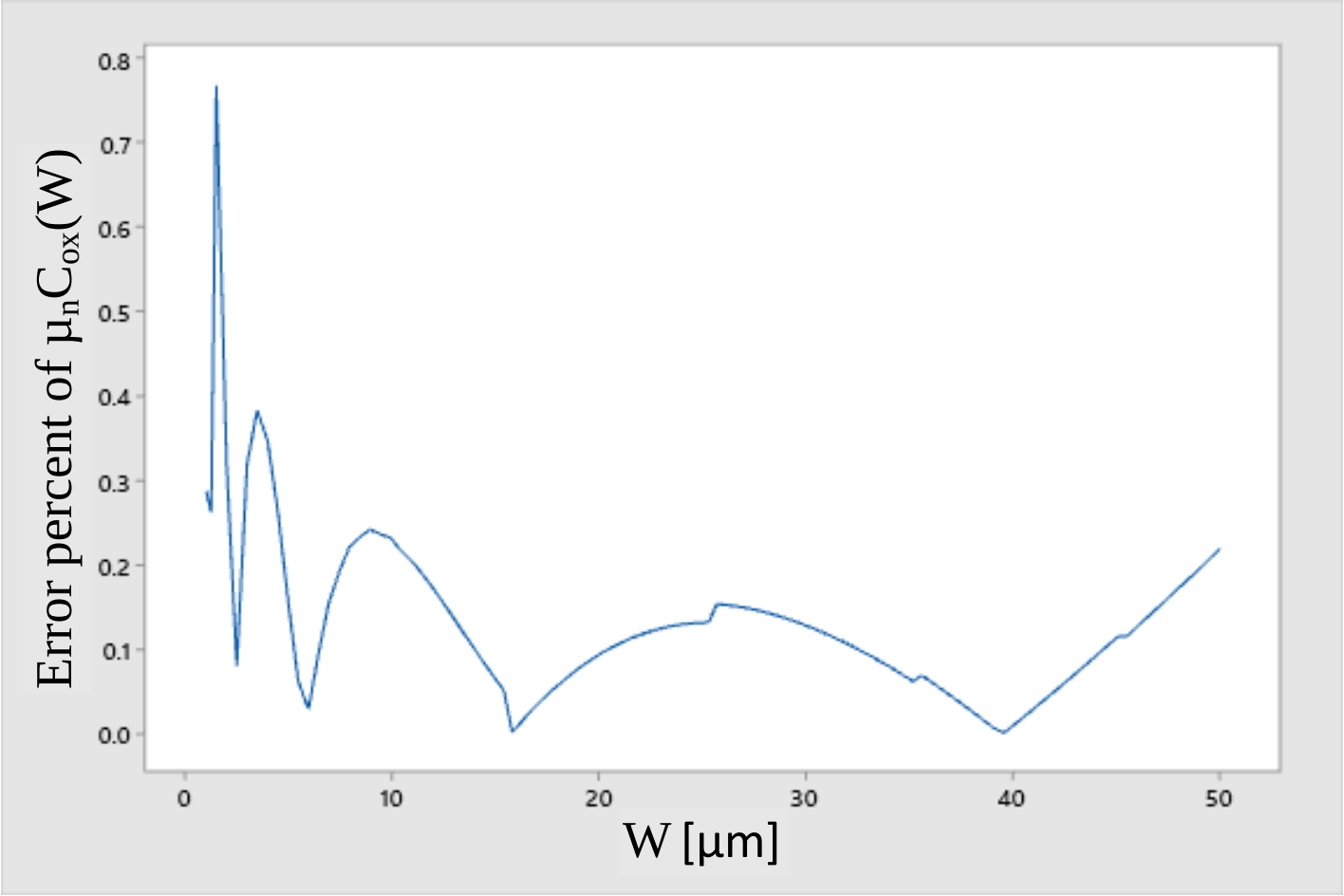

Figure 4.7 Error percent for $\mu_n C_{ox}$ versus W between simulation and regression.

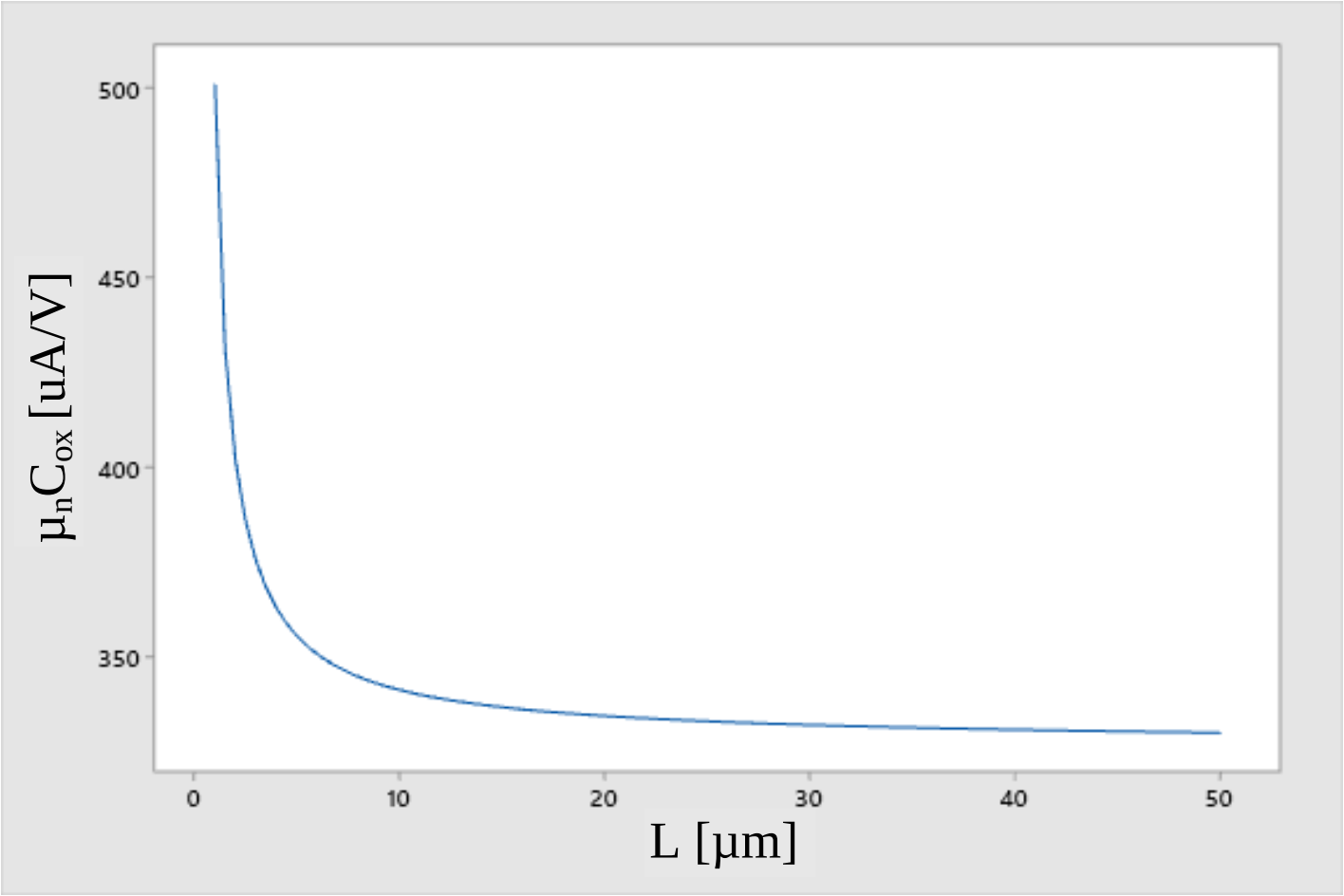

Figure 4.8 $\mu_n C_{ox}$ versus L plot obtained from the circuit simulation while other variables are kept at general values in Table 4.1.

The relationship between $\mu_n C_{ox}$ versus L is a convex asymptotic regression as shown in Figure 4.8. We use the following regression model to estimate this relationship while keeping all the variables at their general values (presented in Table 4.1)

$$\mu_n C_{ox}(L) = \theta_1 + \theta_2 \times \exp\left(-\theta_3 \times L\right) \tag{4.7}$$

After using the nonlinear regression method described in section 4.1.1, θ's are estimated as

$$\mu_n C_{ox}(L) = 333.19 \times 10^{-6} + 185 \times 10^{-6} \times \exp(-404.08 \times L \times 10^{6}) \tag{4.8}$$

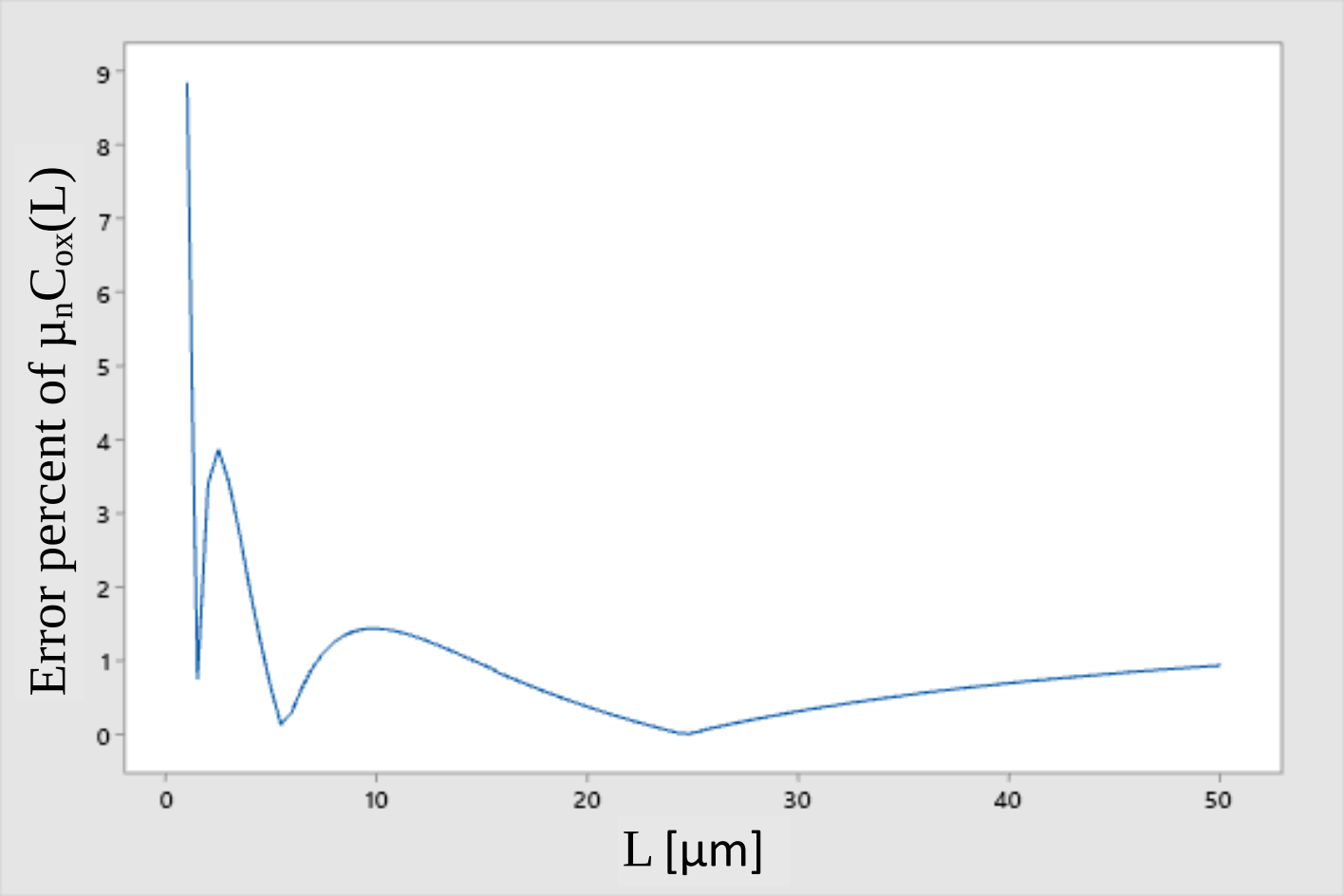

Figure 4.9 Error percent for $\mu_n C_{ox}$ versus L between simulation and regression.

where the error percent plot for the estimated model is shown in Figure 4.9. We can see the error percent is less than 9% and has a decreasing trend.

After finding the individual relationship for $\mu_n C_{ox}$ with W and L, we need to find how W and L jointly change $\mu_n C_{ox}$. We pick ten points starting from 1 μm to 9.91 μm since this range covers most of the variation in the $\mu_n C_{ox}$ based on Figures 4.6 and 4.8. For each point we drive the value of $\mu_n C_{ox}$ from simulation and calculate the value for the $\mu_n C_{ox}(W)$ and $\mu_n C_{ox}(L)$. We pick the following regression model to estimate this relationship

$$\mu_n C_{ox}(W, L) = \beta_0 + \beta_1 \mu_n C_{ox}(W) + \beta_2 \mu_n C_{ox}(L) \tag{4.9}$$

We use linear regression method leveraging the least square error algorithm presented in 4.1.2 to estimate β's values so that $\mu_n C_{ox}(W, L)$ most closely fits the simulation plots

$$\begin{aligned} \mu_n C_{ox}(W, L) = 101.06 \times 10^{-6} + 0.2044 \times \mu_n C_{ox}(W) \\ + 0.41887 \times 10^{-6} \times \mu_n C_{ox}(L) \end{aligned} \tag{4.10}$$

It is worth noting that the true relationship is more complex than linear due to one or more interaction terms between $\mu_n C_{ox}(W)$ and $\mu_n C_{ox}(L)$.

To calculate the value of the $I_P$ in the design process of the transistor circuit model, we need to know the value of $r_o$. $r_o$ is so nonlinear with the transistor physical and electrical properties that it is very difficult to come up with a precise regression model. However, $r_o$ has a small contribution in sizing because passive current $I_p$ generally forms very small portion of the drain current in saturation region as we can see from Figures 3.1 and 3.2. Moreover, we know that $r_o$ is proportional with L. Therefore, we use a very simple regression model of $r_o$ versus L. To do this, we simulate $r_o$ with different values of L with drain current fixed at 1 μA in saturation region. We pick 1 μA because it is an average current that we will use in our design in the next chapter. Then we use linear regression to estimate this relation as shown in Figure 4.10. The regression model for this relation is

$$r_o = 8.64 \times 10^6 + 1.4014 \times 10^{12} \times L \quad (4.11)$$

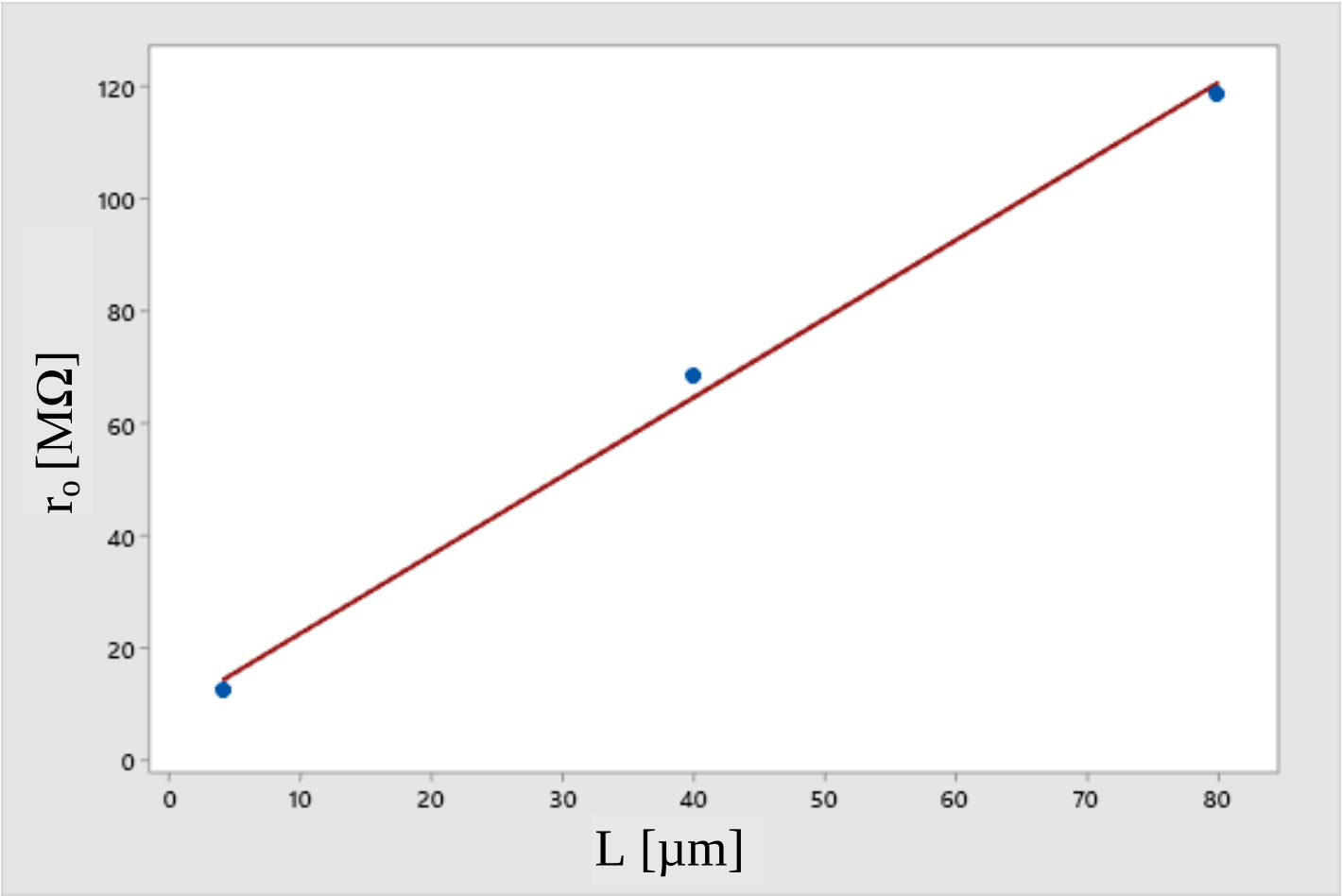

Figure 4.10 $r_o$ versus L plot obtained from the circuit simulation while other variables are kept at general values in Table 4.1.

4.2.1 PMOS

To model the PMOS characteristic, we use the same approach as NMOS. We use the Figure 4.11 setup to derive the plots that shows the relationship for $\mu_n C_{ox}$ and $V_{th}$ with W and L. We use Cadence Spectra Circuit Simulator to obtain the plots.

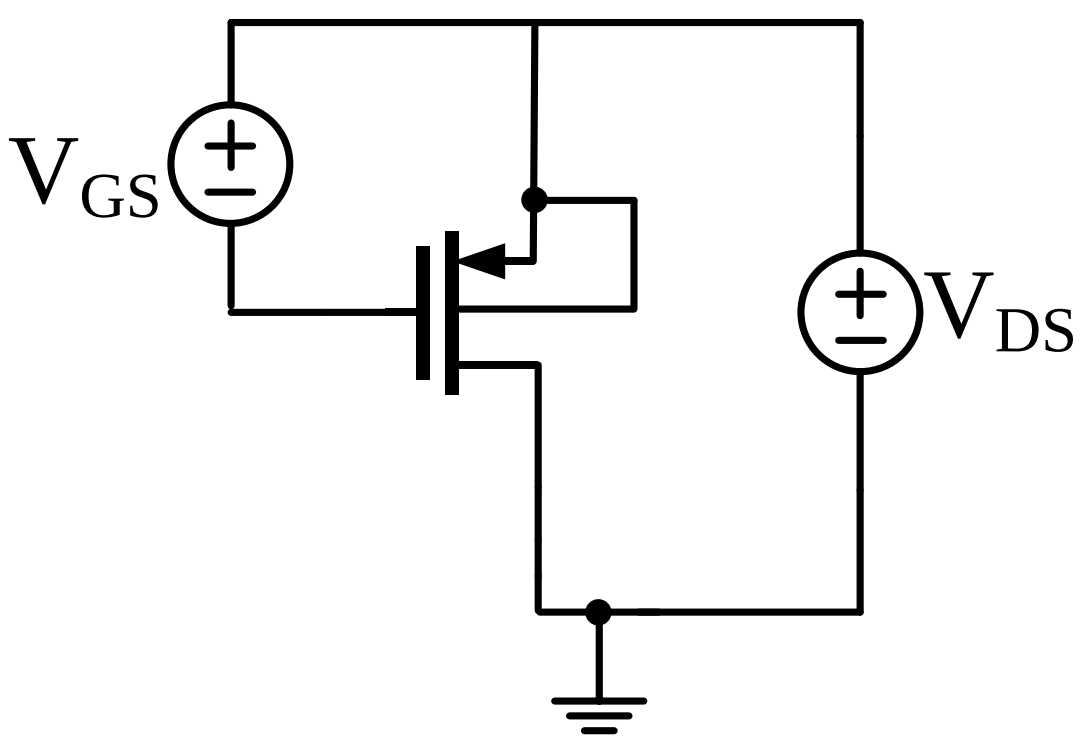

Figure 4.11 Simulation setup for obtaining PMOS $\mu_n C_{ox}$ and $V_{th}$ plots.

By leveraging the nonlinear regression method described in 4.1.1, We find the following estimated models for $\mu_n C_{ox}$ versus W and $\mu_n C_{ox}$ versus L using the same approach as NMOS.

$$\mu_n C_{ox}(W) = 89.038 \times 10^{-6} - 19.262 \times 10^{-6} \times \exp(-0.34468 \times W \times 10^{6}) + 0.09278 \times W \quad (4.12)$$

$$\mu_n C_{ox}(L) = 68.21 \times 10^{-6} + 9.469 \times 10^{-6} \times \exp(-0.50698 \times L) \quad (4.13)$$

Then we find $\mu_n C_{ox}$ versus W and L by using the linear regression.

$$\mu_n C_{ox}(W, L) = 26.07 \times 10^{-6} + 0.6815 \times \mu_n C_{ox}(W) + 0.6684 \times \mu_n C_{ox}(L) \quad (4.30)$$

To estimate the relation between $V_{th}$ verses W and L, we use the Michaelis-Menten regression model which has a similar trajectory as the asymptotic regression model but is more accurate when there is less variation between two variables. $V_{th}$ in the PMOS varies less when sweeping W or L than $V_{th}$ in the NMOS which makes Michaelis-Menten a better choice to estimate PMOS $V_{th}$ versus W or L. The regression model for $V_{th}$ verses W is

$$V_{th}(W) = \frac{\theta_1 \times W}{\theta_2 + W} \quad (4.31)$$

After applying the nonlinear regression method presented in 4.1.1, the estimated model is

$$V_{th}(W) = \frac{-414.01e^3 \times W \times 10^6}{-35.91 \times 10^{-3} - W \times 10^6} \quad (4.32)$$

Similarly, for $V_{th}$ verses L, we have

$$V_{th}(L) = \frac{-396.94 \times 10^3 \times L \times 10^6}{-14.34 \times 10^{-3} + L \times 10^6} \quad (4.33)$$

For the modeling of the threshold voltage verses W and L combined, we use the same approach as the NMOS.

$$V_{th}(W, L) = 1.445 + 2.81 \times V_{th}(L) + 1.79 \times V_{th}(W) \quad (4.34)$$

To predict $r_o$, we simulate $r_o$ with different values of L with drain currents fixed in 1 μA in saturation region like the NMOS. We derive the following regression model to predict PMOS $r_o$

$$r_o = 55.9 \times 10^6 + 6.712 \times 10^{12} \times L \quad (4.35)$$

# CHAPTER V

# DESIGNING A CFIA USING THE PROPOSED METHOD

In this chapter, we use the proposed method to design a Current Feedback Instrumentation Amplifier (CFIA). First, an overview of the CFIA concept is presented. Then small signal analysis of the selected CFIA is reviewed. Subsequently, we show the procedure to size transistors in the circuit. At the end, we present the results and compare the accuracy of the proposed method with $g_m/I_D$ and brute force.

## 5.1 CFIA

The need to amplify differential weak signal with high common mode is one of the main goals of many analog circuits. One of the solutions is using CFIA with high Common Mode Rejection Ratio CMRR. Features like low power consumption and low cost make CFIA a popular choice for analog circuit designers. Figure 5.1 shows the basic block diagram of CFIA. Two unity gain buffers in the input guarantee a high input impedance [15]. For this circuit, the current in resistor $R_g$ is as

$$i_g = \frac{v_1 - v_2}{R_g} \tag{5.1}$$

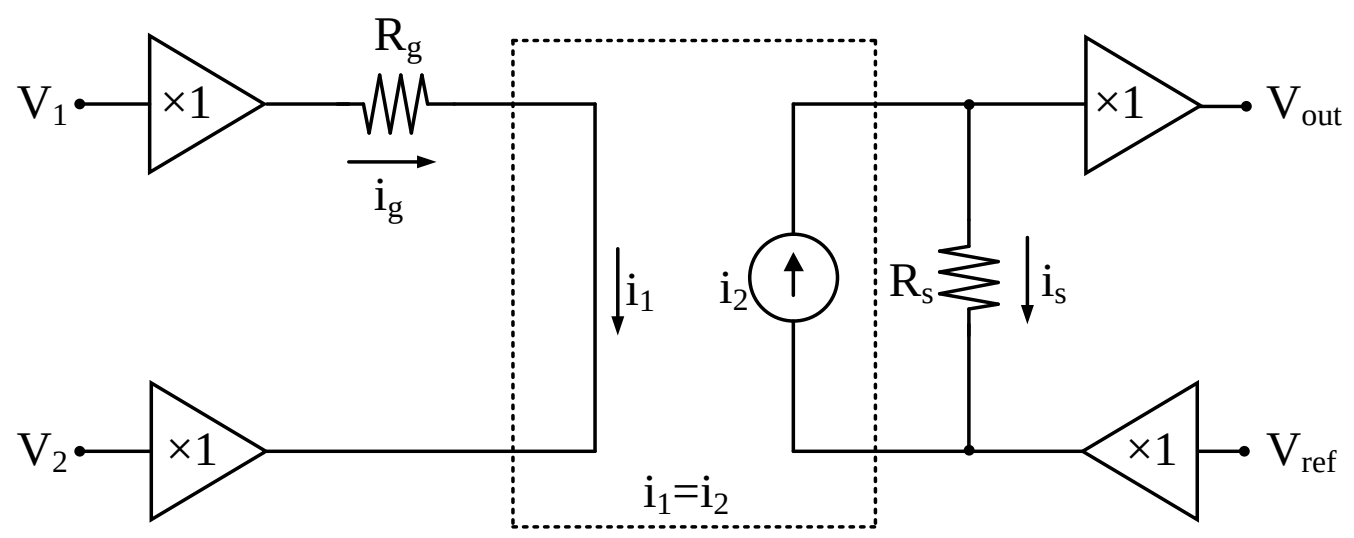

Figure 5.1 Block diagram of CFIA.

With a KVL on the output loop, the output voltage is

$$V_{out} = R_s i_s + V_{ref} \tag{5.2}$$

The input circuit works as a transconductance amplifier while the output circuit works as a transimpedance amplifier. So, if the input current and output current are equal, we have

$$V_{out} = \frac{R_s}{R_g}(V_1 - V_2) + V_{ref} \tag{5.3}$$

5.2 Small Signal Analysis

In this section we want to show small signal analysis of the CFIA circuit proposed in [16]. The schematic of the circuit is shown in Figure 5.2.

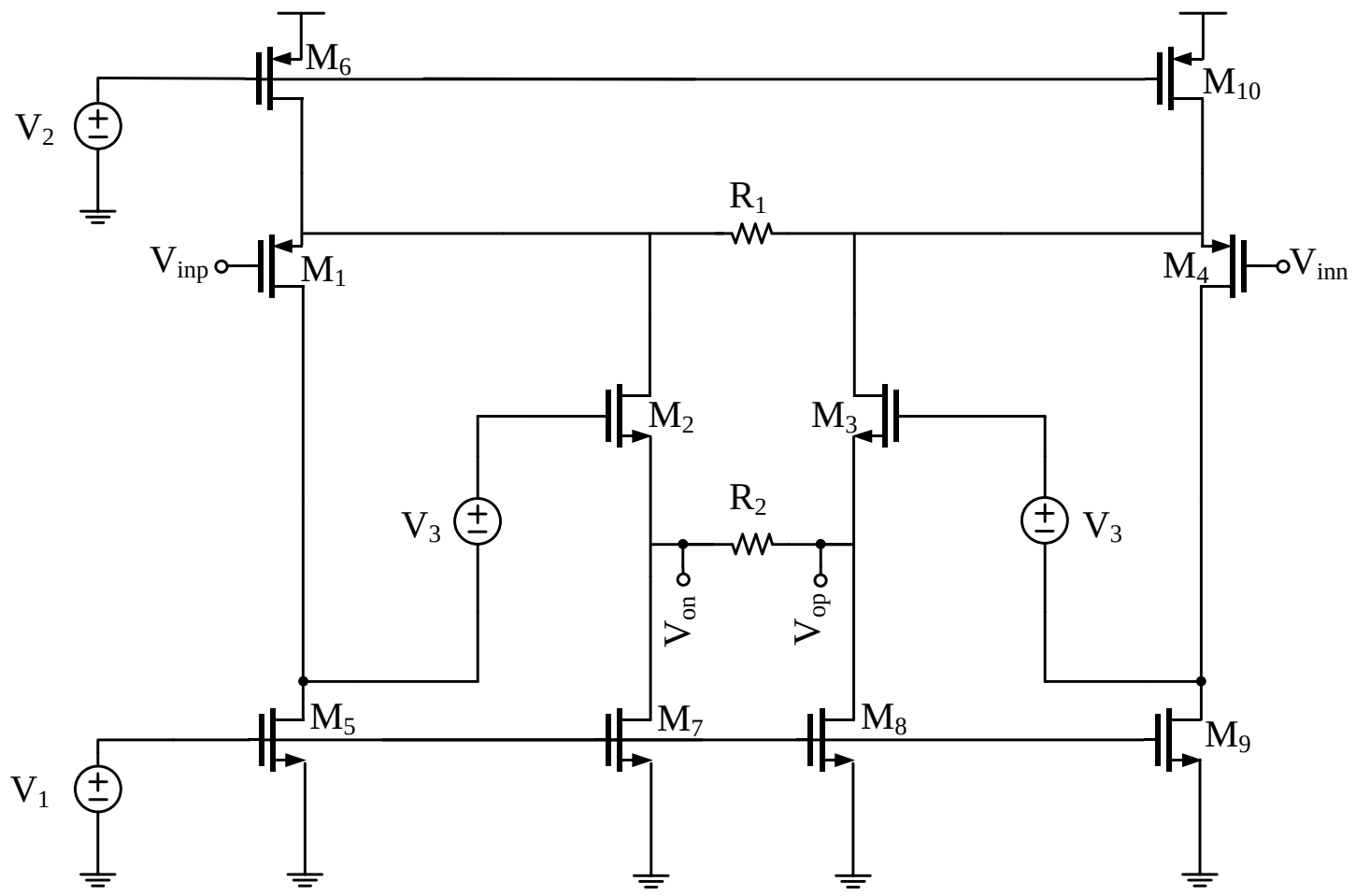

Figure 5.2 CFIA schematic

Since the circuit is symmetric, we only show the small signal analysis for the half circuit as shown in Figure 5.3. For this circuit, we can derive the following equations by writing KVL and KCL equations. By writing a KCL in $D_2$

$$i = i_1 + i_2 + i_3 = \frac{-V_{D2}}{R_1/2} \tag{5.4}$$

With a KCL at $D_1$, we have

$$i_1 = g_{m1}(V_{D2} - V_{in}) + \frac{V_{D2} - V_{G2}}{r_{o1}} \tag{5.5}$$

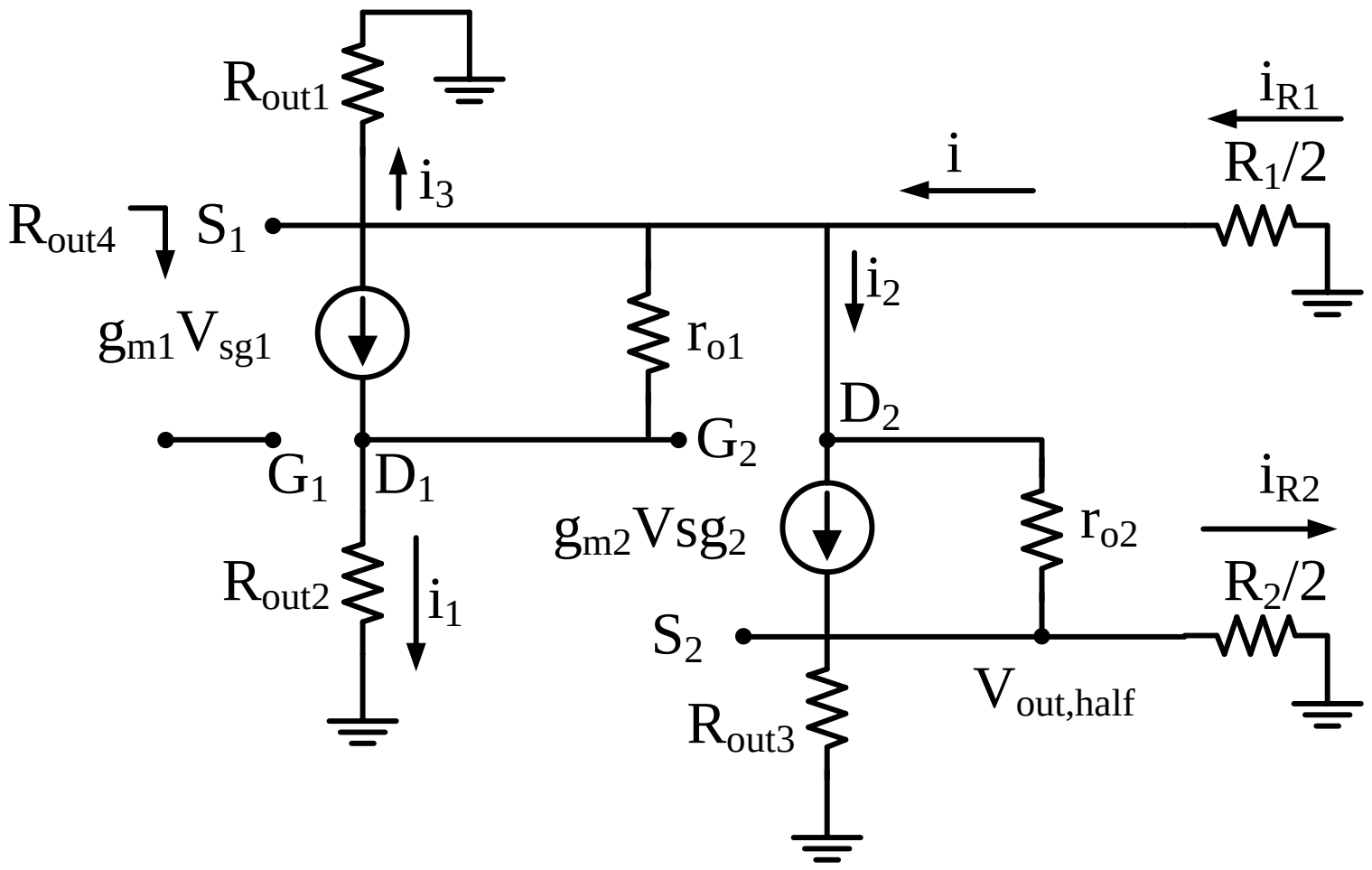

Figure 5.3 Small signal half circuit of the CFIA.

By another KCL at $D_2$

$$i_2 = g_{m2}(V_{G2} - V_{out}) + \frac{V_{D2} - V_{out}}{r_{o2}} \tag{5.6}$$

With a KVL in the loop of $S_1$ to ground, we have

$$i_3 = \frac{V_{D2}}{R_{out1}} \tag{5.7}$$

With a KVL in the loop of $G_2$, $D_1$, and ground, we have

$$V_{G2} = i_1 . R_{out2} \tag{5.8}$$

And for the output voltage

$$V_{out} = i_2 . R_{out3}||R_2/2 \tag{5.9}$$

We can define α as following to make the calculations simple

$$\alpha = \frac{R_{out2}}{1 + R_{out2}/r_{o1}} = R_{out2}||r_{o1} \tag{5.10}$$

We can use (5.5), (5.8) and (5.10) to derive following equation for $i_1$

$$i_1 = \frac{\alpha}{R_{out2}} (g_{m1}(V_{D2} - V_{in}) + \frac{V_{D2}}{r_{o1}}) \tag{5.11}$$

By using (5.6), (5.8) and (5.11), we can derive the following equation for $i_2$

$$\begin{aligned} i_2 = V_{D2}\left(\alpha g_{m2}\left(g_{m1} + \frac{1}{r_{o1}} + \frac{1}{\alpha g_{m2} r_{o2}}\right)\right) \\ -\alpha g_{m1} g_{m2} V_{in} - V_{out}\left(g_{m2} + \frac{1}{r_{o2}}\right) \end{aligned} \tag{5.12}$$

By using (5.4), (5.7), (5.11), and (5.12), we have

$$V_{D2} = \left(V_{in}\left(\frac{\alpha g_{m1}}{R_{out2}/2} + \alpha g_{m1} g_{m2}\right) + V_{out}\left(g_{m2} + \frac{1}{r_{o2}}\right)\right)\frac{1}{\beta} \tag{5.13}$$

where

$$\begin{aligned} \beta = \left(\frac{1}{R_1} + \frac{\alpha}{R_{out2}}\left(g_{m1} + \frac{1}{r_{o1}}\right) + \alpha g_{m1} g_{m2} + \frac{\alpha g_{m2}}{r_{o1}} + \frac{1}{r_{o2}} \right. \\ \left. + \frac{1}{R_{out1}}\right) \end{aligned} \tag{5.14}$$

Therefore, from (5.9), we can obtain the gain expression as

$$\frac{V_{out}}{V_{in}} =$$

$$\frac{\alpha\left(\frac{\alpha g_{m1}}{R_{out2}/2} + g_{m1}g_{m2}\right)\left(\alpha g_{m2}\left(g_{m1} + \frac{1}{r_{o1}}\right) + \frac{1}{r_{o2}}\right) - \alpha\beta g_{m1}g_{m2}}{\beta - (\left(g_{m2} + \frac{1}{r_{o2}}\right)(\alpha g_{m2}\left(g_{m1} + \frac{1}{r_{o1}}\right) + \frac{1}{r_{o2}}) - g_{m2} - \frac{1}{r_{o2}})(R_{out3}||R_2/2)} \tag{5.15}$$

$$(R_{out3}||\frac{R_2}{2})$$

To have the circuit working properly, we need to have $i_{R1} = i_{R2}$. Therefore, the following requirements need to be satisfied

$$R_{out3} \gg R_2 \tag{5.16}$$

$$R_{out1} \gg 1 \tag{5.17}$$

$$r_{o1} \gg 1 \tag{5.18}$$

$R_{out4}$ can be written as

$$R_{out4} = (\frac{1}{g_{m1}}||\frac{1}{g_{mb1}}||r_{o1})(1 + \frac{R_{out2}}{r_{o1}}) \tag{5.19}$$

It is important that the $R_{out4} << R_{out1}$ so that most of the i goes to $i_1$. If we assume

$$r_{o1} \gg R_{out2} \tag{5.20}$$

$R_{out4}$ and α can be simplified as

$$R_{out4} = \frac{1}{g_{m1}} \tag{5.21}$$

$$\alpha = R_{out2} \tag{5.22}$$

Based on the (5.17), (5.18), (5.22), and (5.23), we can simplify β

$$\beta \approx \frac{1}{R_1} + \alpha g_{m1}g_{m2} \tag{5.23}$$

If we assume

$$r_{o2} \gg R_{out2} \tag{5.24}$$

Final expression can be simplified as

$$\frac{V_{out}}{V_{in}} \approx \frac{(R_{out2}g_{m1}g_{m2})^2 - \beta R_{out2}g_{m1}g_{m2}}{\beta} R_2 \sim \frac{-R_{out2}g_{m1}g_{m2}R_2}{1 + R_{out2}g_{m1}g_{m2}R_1} \sim -\frac{R_2}{R_1} \tag{5.25}$$

Finally, based on assumptions in equations (5.16), (5.17), (5.20), and (5.24), the gain can be approximated as $-\frac{R_2}{R_1}$.

## 5.3 Transistor Sizing

In the circuit shown in Figure 5.2, sizing of $M_1, M_5, M_6,$ and $M_7$ will be determined using the transistor circuit model so that it satisfies the CFIA design constraints with proper operation condition. Since the circuit is symmetric, we only show the design of the left half of the circuit. Based on the small signal analysis, we have the following constraints in the circuit to get a gain close to what is achieved in equation (5.25)

$$R_2 \ll r_{o7} \tag{5.26}$$

$$\frac{1}{g_{m1}} \ll r_{o6} \tag{5.27}$$

$$r_{o1} \gg r_{o5} \tag{5.28}$$

$$r_{o2} \gg r_{o5} \tag{5.29}$$

The lengths are set based on the following relationship to satisfy the above constraints

$$r_o = \frac{1}{g_{ds}} = \frac{1}{\lambda . I_{ds}} \propto \frac{L}{I} \quad (5.30)$$

Table 5.1 shows the values for the $L's$. we also have a 15 μW power constraint. Therefore, we select bias current of 1.6 μA for $M_6$, 0.6 μA for $M_1$ and $M_5$, and 1 μA for $M_2$ and $M_7$ to have a power of 5.76 μW which is well below the constraint. Since we trained the regression models for $V_{GS}$ and $V_{DS}$ of 0.5 V and 0.6 V respectively, we use these desired values in sizing process. The desired values are shown in Table 5.1

Table 5.1 Desired transistors values for sizing CFIA.

| Transistor parameters | Transistor name | | | |
|---|---|---|---|---|
| | $M_1$ | $M_5$ | $M_6$ | $M_7$ |
| I [μA] | 0.6 | 0.6 | 1.6 | 1 |
| L [μm] | 40 | 4 | 25 | 80 |
| $V_{GS}$ [V] | -0.5 | 0.5 | -0.5 | 0.5 |
| $V_{DS}$ [V] | -0.6 | 0.6 | -0.6 | 0.6 |

Now, we show the procedure for obtaining W's of transistors. Since the procedure is same for NMOS and PMOS, we only describe setting W of the NMOS. We use equations (3.1), (3.2), (3.3), (4.17), (4.24), and (4.25) based on the flowchart shown in Figure 5.4 to find W, $\mu_n C_{ox}$ and $V_{th}$. First, we set desired values for $I_D$, $V_{DS}$, and $V_{GS}$ based on the small signal analysis. Then $r_o$ can be derived using equation (4.25). Subsequently, W can be predicted using equations (3.2), (3.3), (4.17), and (4.24). We tested the algorithms with different range of W to verify that it always results in a real value for W.

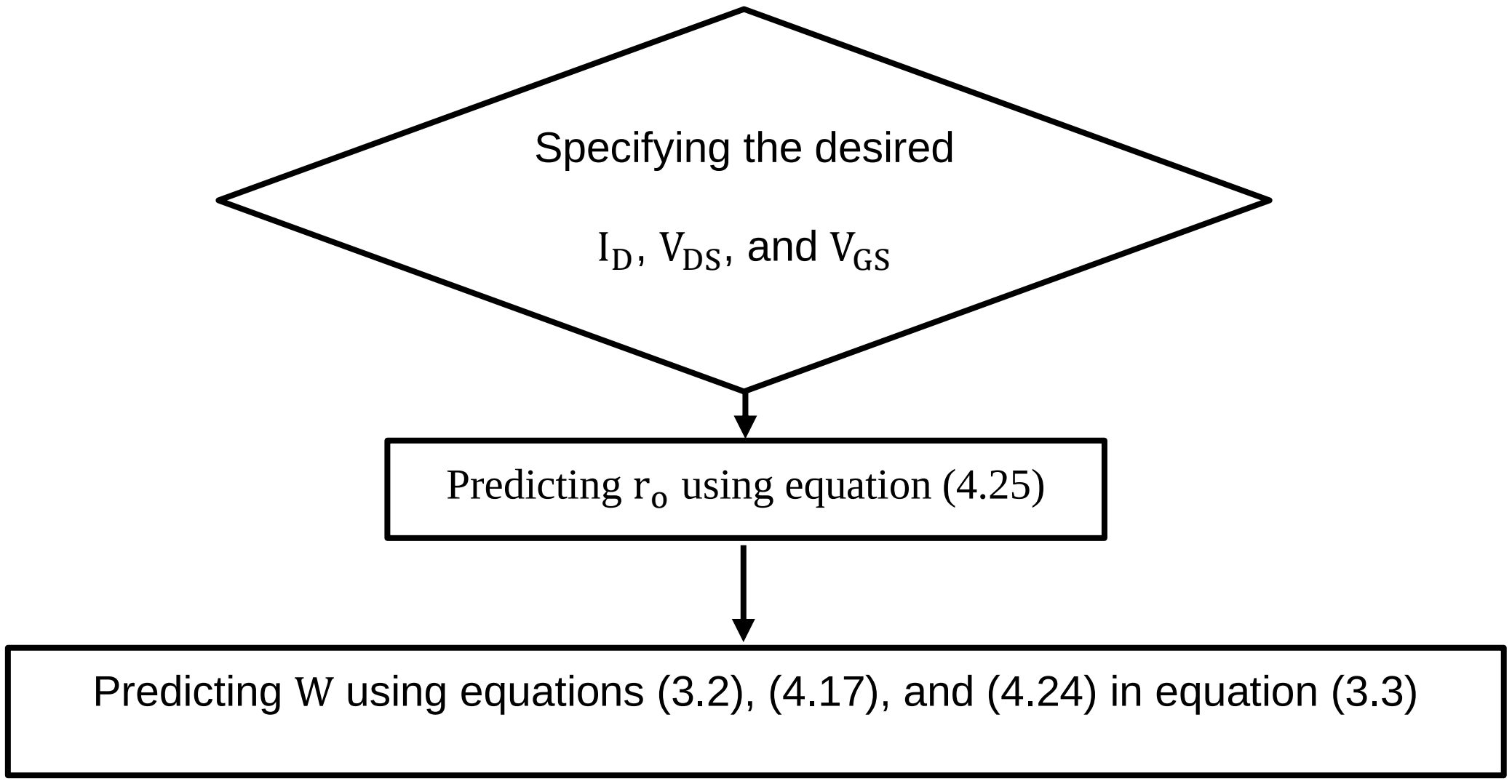

Figure 5.4 The proposed method to predict W of a transistor.

Table 5.2 shows the sizing result using the proposed method. In this table, we use the desired operating conditions and L to predict the value of $r_o$, $V_{th}$, $\mu_n C_{ox}$, and finally W using the procedure described in Figure 5.4. Then, we use cadence to simulate each transistor individually using setups in Figures 4.1 to find $r_o$, $V_{th}$, $\mu_n C_{ox}$, and I values for the predicted W. From Table 5.3, we can see that $V_{th}$ has less than 0.2% error. Also, the error in predicting $\mu_n C_{ox}$ does not grow over 7%. The reason that $\mu_n C_{ox}$ has greater error than $V_{th}$ is when we were modeling $V_{th}$ in the Chapter IV, we had smaller error percent than $\mu_n C_{ox}$. $r_o$ has a bigger error than $\mu_n C_{ox}$ and $V_{th}$ as it is expected from the Chapter V. Since $I_p$ only forms a small portion of the overall current, Inaccurate $r_o$ does not affect overall accuracy of the method noticeably (Figure 3.1).

Table 5.2 Sizing result using the improved transistor circuit model.

| Conditions | Transistor parameter | Transistor names | | | | |
|---|---|---|---|---|---|---|
| | | $M_1$ | $M_2$ | $M_5$ | $M_6$ | $M_7$ |
| Desired operating condition and L | I [μA] | 0.6 | 1 | 0.6 | 1.6 | 1 |
| | $V_{DS}$ [V] | 0.6 | 0.6 | 0.6 | 0.6 | 0.6 |
| | $V_{GS}$ [V] | 0.5 | 0.8 | 0.5 | 0.5 | 0.5 |
| | $V_{SB}$ [V] | 0 | 0.6 | 0 | 0 | 0 |
| | L [μm] | 40 | 20 | 4 | 25 | 80 |
| Predicted | $r_o$ [MΩ] | 324 | 36.6 | 14.2 | 223 | 120 |
| | W [μm] | 73.2 | --- | 0.650 | 119.2 | 22.6 |
| | $\mu_n C_{ox}$ [μA/V] | 84 | --- | 324 | 86.9 | 321 |
| | $V_{th}$ [V] | 0.411 | --- | 0.350 | 0.412 | 0.351 |
| Simulation Result | $r_o$ [MΩ] | 606 | 47.4 | 23.8 | 166.6 | 108.6 |
| | I [μA] | 0.588 | --- | 0.583 | 1.53 | 1.1 |
| | $\mu_n C_{ox}$ [μA/V] | 83 | --- | 304 | 83.5 | 345 |
| | $V_{th}$ [V] | 0.411 | --- | 0.351 | 0.412 | 0.35 |

Table 5.3 Error percent in prediction $\mu_n C_{ox}$ and $V_{th}$.

| Transistor parameters | Transistor error % | | | |
|---|---|---|---|---|
| | $M_1$ | $M_5$ | $M_6$ | $M_7$ |
| $U_n C_{ox}$ | 1.2 | 6.5 | 4 | 6.9 |
| $V_{th}$ | 0 | 0.2 | 0 | 0.2 |
| $r_o$ | 46.5 | 40.3 | 33.8 | 10.2 |

The relation between the $I_D$ versus W for transistors $M_1$, $M_5$, $M_6$, and $M_7$ using the proposed method is shown in Figure 5.4. We design the CFIA using predicted values in the Table 5.2 and measure the $V_{DS}$ values. We can see from Table 5.4 that there is a deviation between expected $V_{DS}$ and actual $V_{DS}$. To understand what causes the deviation, we also obtain the error % of $I_P$, $I_A$, and $V_{DS}$ as it is shown in Table 5.4. Please note, to obtain error % of $I_P$ and $I_A$, circuit 4.1 and 4.11 is used and the $I_P$ and $I_A$ of predicted and desired W is compared to find the error percent.

Then, we run Pearson correlation between L, $I_D$, and error % of $I_P$, $I_A$, and $V_{DS}$. The result is shown in Table 5.5. Only the P-Value between $I_D$ and the error % of $V_{DS}$ shows an evidence that there is a correlation between $I_D$ and the error % of $V_{DS}$ for a significance level of 0.05. That means $I_D$ is the cause of inconsistent $V_{DS}$ deviation in 4 transistors. The coefficient for this correlation is -0.95 which means as $I_D$ increases, error % of $V_{DS}$ decreases.

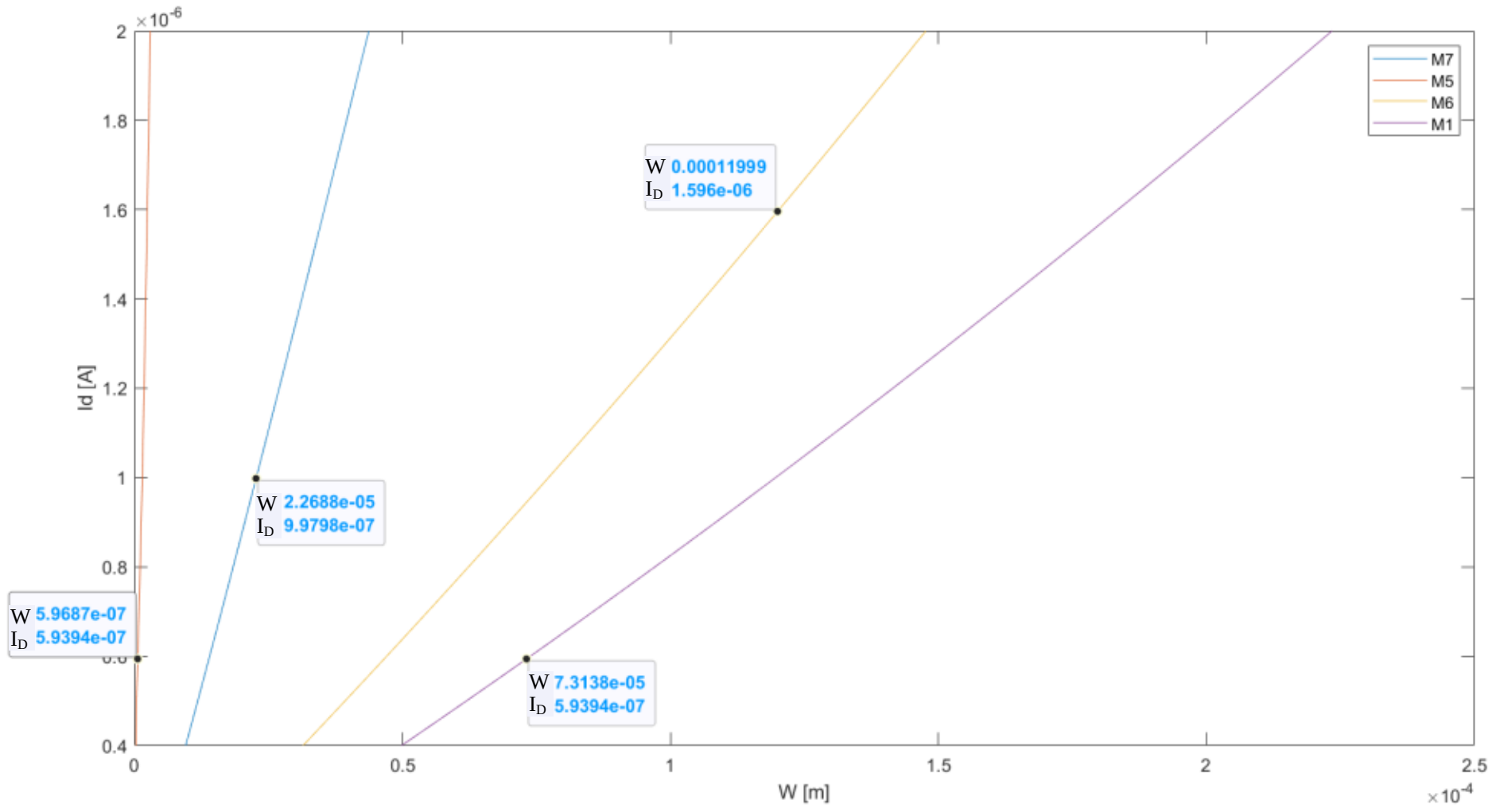

Figure 5.5 $I_D$ versus W for transistors $M_1$, $M_5$, $M_6$, and $M_7$.

Table 5.4 Expected and actual value of $V_{DS}$, $I_P$, and $I_A$ in the designing process.

| Condition | Transistor values | | | |
|---|---|---|---|---|
| | $M_1$ | $M_5$ | $M_6$ | $M_7$ |
| Expected $V_{DS}$ [V] | 0.6 | 0.6 | 0.6 | 0.6 |
| Actual $V_{DS}$ [V] | 1.08 | 0.092 | 0.618 | 0.161 |
| Error % | 80 | 84 | 0.3 | 73 |
| Expected $I_P$ [nA] | 14 | 30 | 39 | 18 |
| Actual $I_P$ [nA] | 11 | 30 | 37 | 20 |
| Error % | 21 | ~0 | 5.1 | 11.1 |
| Expected $I_A$ [uA] | 0.586 | 0.570 | 1.56 | 0.981 |
| Actual $I_A$ [uA] | 0.577 | 0.552 | 1.50 | 1.080 |
| Error % | 1.5 | 3.1 | 3.8 | 10.1 |

Table 5.5 Pearson Correlations.

| Sample 1 | Sample 2 | N | Correlation | 95% CI for ρ | P-Value |
|---|---|---|---|---|---|
| Error % $I_P$ | Error % $V_{DS}$ | 4 | 0.266 | (-0.934, 0.977) | 0.734 |
| Error % $I_A$ | Error % $V_{DS}$ | 4 | 0.048 | (-0.957, 0.965) | 0.952 |
| L | Error % $V_{DS}$ | 4 | 0.143 | (-0.948, 0.971) | 0.857 |
| $I_D$ | Error % $V_{DS}$ | 4 | -0.954 | (-0.999, 0.089) | 0.046 |

To have the circuit working properly, we need to adjust $V_{DS}$ of transistors. We use the transistor circuit model described in the Chapter 3 to adjust the $V_{DS}$. Since $|V_{DS5}| < 0.6$ V, we can increase the $I_p$ of $M_5$ by decreasing the active current of $M_5$ which can be achieved by decreasing W based on Figure 3.1. For the $W_{M5} = 320$ nm, we can obtain $V_{DS5} = 0.584$ V. After adjusting $V_{DS5}$, the $V_{DS1}$ decreases and $V_{DS7}$ increases close to its expected value and circuit works and gives a high absolute gain of 8.17. The adjusted VDS values are shown in Table 5.6. We picked 100 KΩ and 10 KΩ for $R_2$ and $R_1$, respectively. Also, we chose 0.5 V, 1.3 V, and 0.4 V for $V_1$, $V_2$, and $V_3$ respectively.

Table 5.6 Adjusted $V_{DS}$ in designing process of CFIA.

| Condition | Transistor values | | | |
|---|---|---|---|---|
| | $M_1$ | $M_5$ | $M_6$ | $M_7$ |
| Expected $V_{DS}$ [V] | 0.6 | 0.6 | 0.6 | 0.6 |
| Actual $V_{DS}$ [V] | 1.08 | 0.092 | 0.618 | 0.161 |
| Adjusted $V_{DS}$ [V] | 0.593 | 0.584 | 0.622 | 0.572 |

The DC and AC analysis has been shown in Figures (5.5) and (5.6), respectively. DC analysis shows that bias point has a limited swing of 5 mV. This is mainly because we had fixed values for the $V_{DS}$ and $V_{GS}$ in our design problem. The AC analysis confirms the gain of 8.17 and shows Gain Bandwidth (GBW) of 6 MHz. The transient response shown in Figure (5.7) confirms the gain and functionality of the circuit. The simulation result for the sized CFIA is shown in Table 5.7.

Table 5.7 Simulation result for the designed CFIA.

| Gain (V/V) | GBW (MHz) | Power (µW) | CMRR |
|---|---|---|---|
| 8.17 | 6 | 5.76 | 35.83 dB |

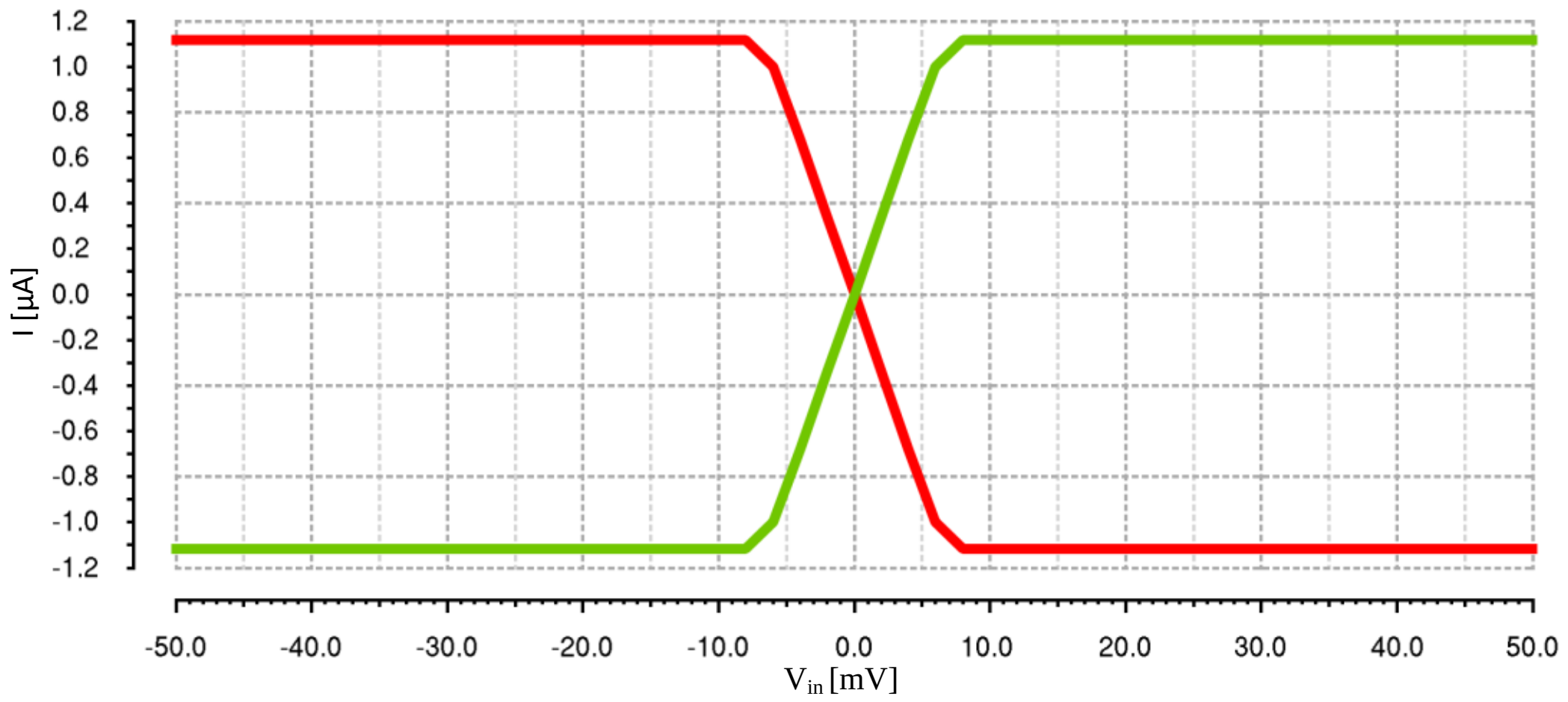

Figure 5.6 DC analysis of the CFIA

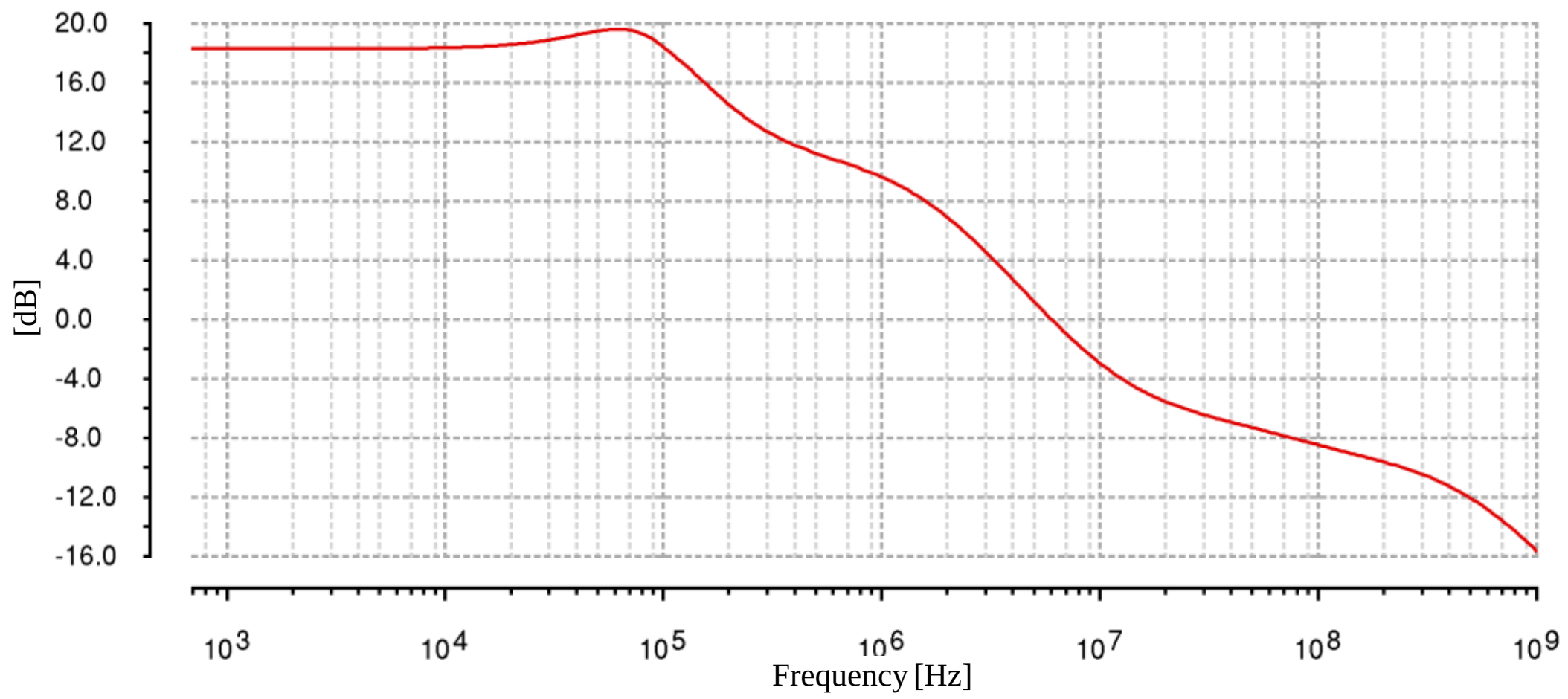

Figure 5.7 AC analysis of the CFIA.

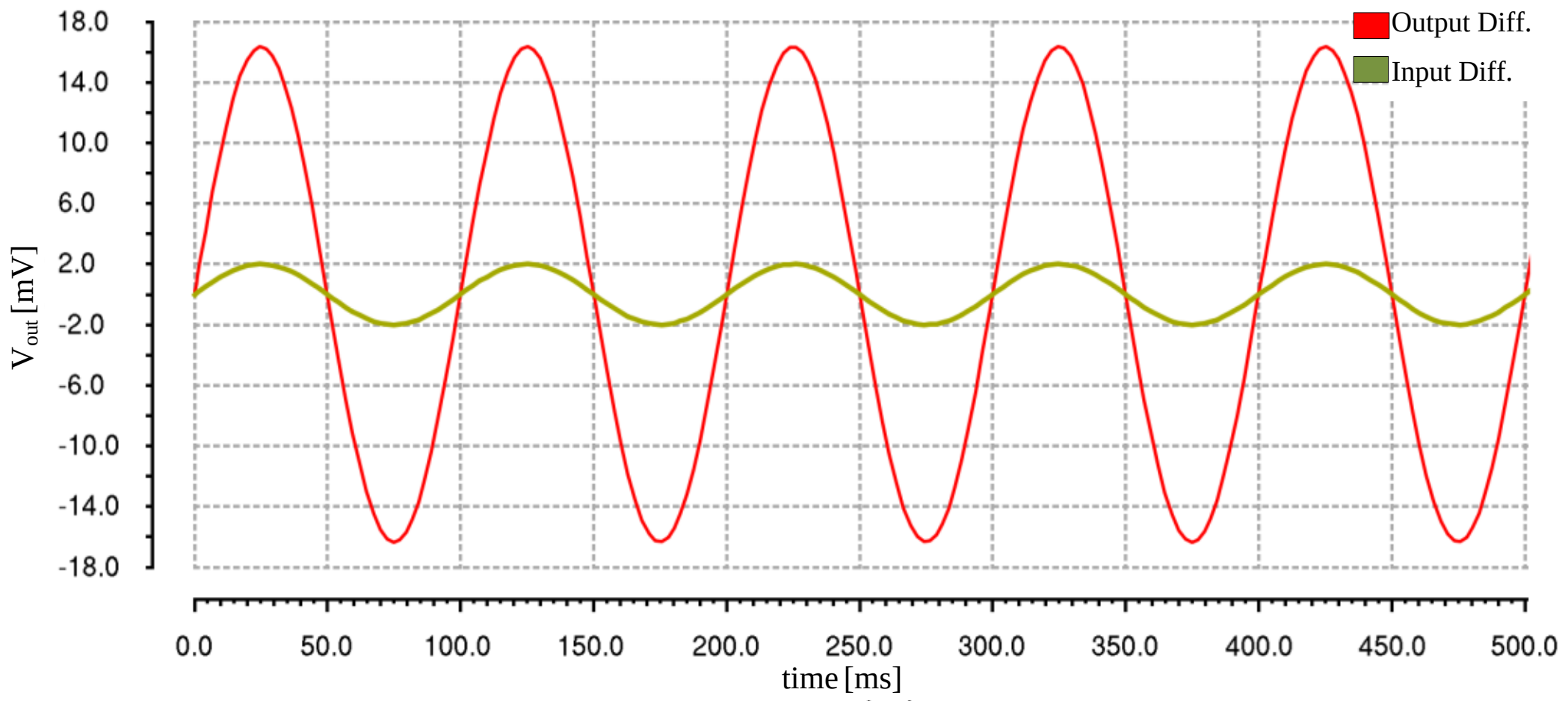

Figure 5.8 Transient analysis of the CFIA.

Table 5.8 shows the corner analysis of the designed circuit. The circuit is sensitive to the thermal and power supply variation. This is mainly due to the limitation in $\mathbf{V}_{DS}$ and $\mathbf{V}_{GS}$ in the design process.

Table 5.8. Corner analysis of designed CFIA

| Corner test | $V_{DS}(mV)$ | | | |
|---|---|---|---|---|
| | M1 | M5 | M6 | M7 |
| VDD=1.6 V, T=125 C | -965 | 22.8 | -985 | 14.7 |
| VDD=1.9 V, T=125 C | -10 | 1850 | -40 | 915 |
| VDD=1.9 V, T=-40 C | -4 | 1863 | -33 | 1068 |
| VDD=1.6 V, T=-40 C | ~0 | ~0 | 1600 | ~0 |

Table 5.9 shows the performance comparison between the proposed method, $g_m/I_D$ and brute-force method. We can see the proposed method is the fastest way to find the W and requires no simulation. It also outperforms $g_m/I_D$ method in accuracy. Bad performance of $g_m/I_D$ in this comparison is because of the focused capacity of this comparison but in general, $g_m/I_D$ offers the most flexibility to satisfy the design specifications. The brute-force method yields the most accurate result due to the employment of direct circuit simulation to obtain the W. However, to have the accuracy equal to 100%, we need many steps in simulator and high reading accuracy which subsequently increase the design time. Please note, these experiments are done with 1000 linear steps in Cadence Spectra with 16 cores at 2300 MHz, and the proposed method is run on Intel Core i7 at 1.8 GHz.

Table 5.9 Performance comparison between the proposed method, $g_m/I_D$ and brute-force.

| Method | Transistor | # Simulations | Time [s] | Predicted W [μm] | Accuracy |
|---|---|---|---|---|---|
| Proposed method | $M_1$ | 0 | 0.7 | 73.2 | 98% |
| | $M_5$ | 0 | 0.7 | 0.650 | 96% |
| | $M_6$ | 0 | 0.7 | 119.2 | 96% |
| | $M_7$ | 0 | 0.7 | 22.6 | 90% |
| $g_m/I_D$ | $M_1$ | 1 | 51 | 66.51 | 89.1% |
| | $M_5$ | 1 | 49 | 0.743 | 90% |
| | $M_6$ | 1 | 38 | 110.72 | 89.2% |
| | $M_7$ | 1 | 34 | 22.77 | 89.6% |
| Brute-Force | $M_1$ | 1 | 30 | 74.61 | ~100% |
| | $M_5$ | 1 | 44 | 0.677 | ~100% |
| | $M_6$ | 1 | 45 | 124.12 | ~100% |
| | $M_7$ | 1 | 25 | 20.64 | ~100% |

# CHAPTER VI

# CONCLUSION

In this paper, a fast method to design an analog circuit for deep sub-micron CMOS fabrication processes is proposed. Regression algorithms are used to improve the transistor circuit model equations to eliminate the need for simulation software for sizing each transistor. The proposed method reduces the time for sizing a transistor more than 98% and 97% compared to $g_m/I_D$ and brute-force method, respectively. We illustrated the efficiency of the proposed method through designing a CFIA and we showed that the model achieves a high accuracy of over 90% in predicting the size of a transistor. The proposed method is an interactive and flexible approach as it offers the freedom to systematically adjust any transistor in the circuit to achieve the desired objective.

In future, we will include the effect of $V_{SB}$, $V_{GS}$, and $V_{DS}$ in changing $\mu_n C_{ox}$ and $V_{th}$ to generalize the proposed model for every design. We consider integrating reinforcement learning to find the optimum desired DC operating point to minimize the design dependence on the designer knowledge to subsequently make the design process automatic.